\documentclass{aa}

\usepackage{graphicx}
\usepackage{placeins}
\usepackage{amsmath}
\usepackage{txfonts}
\usepackage{orcidlink}
\hypersetup{hidelinks}

\newcommand{\BJaaa}{56.8}
\newcommand{\BJaaaErr}{1.5}
\newcommand{\TJaaa}{0.51 \pm 0.04}
\newcommand{\thJaaa}{55 \pm 1}
\newcommand{\lamJaaa}{6.0 \pm 0.2}
\newcommand{\chiredJaaa}{5.0}
\newcommand{\BJbbb}{64.4}
\newcommand{\BJbbbErr}{1.7}
\newcommand{\TJbbb}{4.2 \pm 0.6}
\newcommand{\thJbbb}{47 \pm 2}
\newcommand{\lamJbbb}{4.5 \pm 0.2}
\newcommand{\chiredJbbb}{8.6}
\newcommand{\BJddd}{51.1}
\newcommand{\BJdddErr}{1.4}
\newcommand{\chiredJddd}{2.2}
\newcommand{\PJaaa}{105.349}
\newcommand{\PJddd}{143.765}
\newcommand{\LcycJaaa}{4.2 \times 10^{30}}
\newcommand{\fareaJaaa}{8.9 \times 10^{-4}}
\newcommand{\MdotJaaa}{5.9 \times 10^{-13}}
\newcommand{\muJaaa}{1.9 \times 10^{34}}
\newcommand{\LcycJbbb}{1.9 \times 10^{31}}
\newcommand{\fareaJbbb}{9.6 \times 10^{-5}}
\newcommand{\MdotJbbb}{2.7 \times 10^{-12}}
\newcommand{\muJbbb}{2.2 \times 10^{34}}
\makeatletter
\renewcommand{\aa@acknowledgementsname}{Acknowledgments}
\makeatother
\begin{document}

\title{Magnetic-field constraints for four magnetic cataclysmic
variables from simultaneous continuum and cyclotron modeling}

\titlerunning{Magnetic-field constraints from cyclotron modeling}
\authorrunning{J. Lin et al.}

\author{Jiamao Lin\inst{1,2}\orcidlink{0009-0008-9942-620X}
\and Wenyuan Guo\inst{1,2}\orcidlink{0009-0000-5741-4456}
\and Yongkang Sun\inst{4}\orcidlink{0000-0002-3935-2666}
\and Chengyuan Li\inst{1,2}\orcidlink{0000-0002-3084-5157}
\and Wen Hou\inst{3}\orcidlink{0000-0003-0716-1029}
}

\institute{School of Physics and Astronomy, Sun Yat-sen University,
Zhuhai 519082, China\\
\email{linjm66@mail2.sysu.edu.cn}
\and
CSST Science Center for the Guangdong-Hongkong-Macau Greater Bay Area,
Sun Yat-sen University, Zhuhai 519082, China
\and
CAS Key Laboratory of Optical Astronomy, National Astronomical
Observatories, Chinese Academy of Sciences, Beijing 100101, China
\and
Department of Astronomy, University of Chinese Academy of Sciences,
Beijing 100049, People's Republic of China
}

\date{}

\abstract
{Wide-field spectroscopic surveys are finding magnetic cataclysmic
variable (CV) candidates, but their magnetic fields are rarely measured
because cyclotron humps must be separated from uncertain continua and
assigned to discrete harmonics.}
{We applied a simultaneous continuum--cyclotron analysis to four DESI/LAMOST
magnetic-CV candidates, aiming to determine or bound their fields and
measure their periods.}
{We fit a white dwarf atmospheric model, a hot-spot blackbody, and an
isothermal constant-$\Lambda$ cyclotron component in one inverse-variance
weighted model. Profiling the chi-square goodness of fit over the field
strength exposes competing harmonic assignments, which we report as
multiple acceptable branches when necessary, and posterior sampling gives
branch-conditional uncertainties. We validated the
field recovery on EQ~Cet, BS~Tri, and three SDSS polars with published
fields, searched 6--7 yr of ZTF photometry for periods and aliases, and
used archival UV-to-mid-infrared photometry to bound the
donors.}
{The fit gives one robust field,
$B = \BJaaa\pm\BJaaaErr$\,MG, for J0005$+$2941.  The remaining spectra are
less decisive: J0022$+$1340 has two indistinguishable branches at
$B \simeq 64.4$ and 77.7\,MG; J0749$+$3654 has competitive
solutions near 29.3 and 96.4\,MG; and J0035$+$4333 yields an indicative
single but broad minimum at $B = 51.1\pm1.4$\,MG.
The ZTF data give 105.349\,min for J0005$+$2941, with a narrow
eclipse-like minimum, making it a rare short-period eclipsing-polar
candidate; $80.4062\pm0.0008$\,min for J0749$+$3654 from
double-humped cyclotron beaming; and 143.8\,min for J0035$+$4333.
Line diagnostics show strong He\,II~$\lambda$4686 in J0022$+$1340 and
J0035$+$4333 and flat-to-inverted Balmer decrements in all four,
consistent with magnetically channeled accretion. Archival
mid-infrared photometry excludes donors earlier than about M4.5--M5\,V
in three of the systems and donors earlier than about M3\,V in J0035$+$4333.}
{We thus measure or bound the field of every target and identify
J0005$+$2941 as a rare eclipsing-polar candidate below the period gap.
Joint continuum--cyclotron fitting gives secure fields when the humps are
well resolved and explicit branch constraints otherwise.}


\maketitle
\nolinenumbers

\section{Introduction}
\label{sec:intro}

Polars (AM Herculis stars) are cataclysmic variables (CVs) in which a
strongly magnetized white dwarf (WD; $B$ of $ \sim 10$--240\,MG) accretes from a
late-type donor without forming an accretion disk
\citep[e.g.,][]{1995warner,wickramasinghe2000,hellierbook,knigge2011}.
The magnetic field channels the accretion flow to one or both magnetic
poles, where a standing shock heats the plasma to $\sim$5--50\,keV; the shock cools
through thermal bremsstrahlung and through cyclotron emission, the latter
imprinting broad, harmonically spaced humps on the optical/near-infrared
spectrum \citep{1979Natur.281...47V,1988prco.book....3W}.

The magnetic field strength is the key parameter governing the structure
and evolution of these systems.
Magnetic CVs make up 20\%--36\% of the CV population
\citep{2015ferrario,2020pala,2023inighta}, and the origin of their fields
-- inheritance, binary-interaction dynamos, or field growth during
detached phases -- remains debated
\citep{2008tout,2018briggs,2020belloni,2021schreiber}.
The field also modifies angular momentum loss: the 2--3\,hr period gap
that characterizes nonmagnetic CVs is much less pronounced among
polars \citep{2021schreiber,2024schreiber}, which is attributed to a
reduction in the magnetic braking by the WD field.
Population-level tests of these ideas require homogeneous field
measurements for large samples; the recent PolarCat compilation
\citep{2025schwope} lists 242 polars, but the field strengths therein
derive from heterogeneous methods -- Zeeman splitting in low states,
circular spectropolarimetry, cyclotron harmonic spacing, or Zeeman
tomography -- and from different accretion states.
Wide-field spectroscopic and time-domain surveys are now delivering
magnetic CV candidates in large numbers
\citep{2023inighta,2025desidr1,2023ApJ...945..141R,2025A&A...696A.242V,2025lin,2026hou},
mostly with single-epoch survey spectra: moderate resolving power
($R\sim2000$), signal-to-noise ratios of a few to a few tens per pixel,
and a flux calibration that is homogeneous across the survey but not
tailored to the individual object. Field measurements exist for only a
minority of them.
For most of these systems, spectroscopy of cyclotron emission features is
the most practical field diagnostic available from a single survey
spectrum: the harmonic spacing yields the field of the dominant emission
region even when time-resolved polarimetry is unavailable
\citep{wickramasinghe1987,2020kolbina,2022MNRAS.511...20K}.
This applies to the field of the accretion region at the magnetic pole.
Zeeman features in optical and UV spectra are an equally
effective, complementary diagnostic of the mean longitudinal field at
the photosphere \citep[Fig.~3 of][]{Cunningham2025} but require the
photosphere to be visible, i.e.,\ a low state or a faint period bouncer.

Extracting reliable fields from single-epoch survey spectra is, however,
not straightforward.
The cyclotron humps sit on a composite continuum (WD photosphere,
accretion-heated spot, possibly the donor, and a recombination
continuum), and the inferred cyclotron parameters (especially the
optical depth parameter, $\Lambda,$ and viewing angle, $\theta$) depend
strongly on how that continuum is removed.
The harmonic number ($n$) assigned to an observed hump can also be ambiguous
when only one or two humps fall in the optical window, producing discrete
families of acceptable combinations of field strength ($B$) and harmonic
number ($n$; \citealt{wickramasinghe2000,littlefield2018,2022MNRAS.511...20K,Littlefield2023J1344,2025schwope});
SDSS~J1344$+$2044, for example, admits both $\sim$56 and $\sim$42\,MG
harmonic assignments \citep{Littlefield2023J1344}.
A two-step fit -- a continuum fitted in fixed wavelength windows, then
the residual fitted alone -- is efficient for catalog-scale work, but in
ambiguous spectra the recovered field can depend on the continuum
placement and on the harmonic assignment. Such a fit underlies the
cyclotron fields in our catalog \citep{2026hou} of CVs from Data Release~1 (DR1) of the Dark Energy Spectroscopic
Instrument \citep[DESI;][]{2025desidr1}.
Here we take a different route. We fit the continuum and the cyclotron
component simultaneously in a single weighted likelihood so that
continuum uncertainty propagates directly into the cyclotron parameters,
and we mapped the harmonic ambiguity explicitly by profiling $\chi^2$ over
the field strength. Where more than one harmonic assignment is
statistically acceptable, we quote the alternatives explicitly as separate
branches.
We first validated the method on polars with independently known fields:
the benchmarks EQ~Cet and BS~Tri, plus three polars with archival
spectra from the Sloan Digital Sky Survey (SDSS; \citealt{Ahumada2020}): MQ~Dra, PZ~Vir, and SDSS~J1344$+$2044. The method recovered the published fields
where the humps are well resolved, and we then applied it to the four
targets.

We applied this analysis to four magnetic CV candidates with cyclotron-like
spectral features: three from DESI DR1
and one from the Large Sky Area
Multi-Object Fiber Spectroscopic Telescope
\citep[LAMOST;][]{Cui2012} DR11 (Sect.~\ref{sec:data}).
All three DESI targets were identified as magnetic CVs with cyclotron
features in our DESI DR1 catalog \citep{2026hou}, which also reports
first-pass cyclotron fields of $\approx$45 and 73\,MG for J0005$+$2941 and
J0022$+$1340 from the two-step fit. J0749$+$3654, independently
classified as a polar by \citet{2023inighta} (SDSS~J074917.11$+$365428.0,
MGAB--V3545), had cyclotron features too weak for a field measurement. Here we re-derived these fields with the joint method and mapped their
harmonic ambiguity.
For each source, we combined the spectral analysis with a period search in
the full \textit{Zwicky} Transient Facility (ZTF) archive \citep[][]{Bellm2019,Masci2019}, paying particular attention to alias structure and to accretion
state changes.
Together, these provide one robustly determined field (DESI~J0005$+$2941),
two branch-ambiguous cases reported as multiple acceptable branches
(DESI~J0022$+$1340 and DESI~J0749$+$3654), one indicative broad single
minimum (LAMOST~J0035$+$4333), orbital or beaming periods for three, and,
most significantly, a new eclipsing-polar candidate below the period gap
(DESI~J0005$+$2941).
Section~\ref{sec:data} describes the data, Sect.~\ref{sec:methods} the joint model and the timing
analysis, Sect.~\ref{sec:results} presents per-source results,
Sect.~\ref{sec:discussion} discusses population context and systematic
limitations, and Sect.~\ref{sec:conclusions} summarizes our findings.

\section{Observations and data}
\label{sec:data}

\subsection{Spectroscopy}
\label{sec:data_spec}

The three DESI targets were selected by visual inspection -- for broad,
quasi-periodic humps superposed on a blue continuum, the characteristic
signature of optical cyclotron emission -- from our catalog of
CVs identified in the co-added DESI DR1 spectra
\citep{2026hou}, retaining those whose humps are strong enough for the
harmonic spacing to be measured.
Table~\ref{tab:sample} lists their coordinates, \textit{Gaia} $G$ magnitudes,
geometric distances, and the adopted reddening.
The co-added spectra cover 3600--9800\,\AA{} at a resolving power of
$R \simeq 2000$--5000 and are flux-calibrated in units of
$10^{-17}\,\mathrm{erg\,s^{-1}\,cm^{-2}}\,\mbox{\AA}^{-1}$
\citep{2025desidr1}.
Each is the survey co-add of all exposures of the target in DR1, which
covers the main-survey observations taken between 2021 May and 2022 June
together with the reprocessed Survey Validation data
\citep{2025desidr1}; the median signal-to-noise ratio over the fitted
band is $\simeq6$ per pixel for J0005$+$2941 and $\simeq12$ for the other
two. Because the flux scale is absolute and
the \textit{Gaia} distances are good to 7--14\%, the photospheric solid angle can
be bounded directly (Sect.~\ref{sec:joint}).

LAMOST~J003553.36$+$433341.4 was selected in the same way, from the
LAMOST DR6 CV catalog of \citet{Sun2021}, and was observed twice by the
LAMOST low-resolution survey \citep[3700--9000\,\AA,
$R\sim1800$;][]{Cui2012}, in 2014 October and 2016 November (5400\,s
each).
We analyzed only the 2014 spectrum, whose median $\mathrm{S/N}\simeq5$
per pixel over the fitted band (4--7 in the red arm) exceeds that of the
2016 spectrum (2--5).
Unlike the DESI co-adds, LAMOST spectra are calibrated in relative flux
only, so the absolute scale is left free and no distance prior is
applied; the zero-point-corrected \textit{Gaia} parallax
\citep[$1.01\pm0.43$\,mas;][]{GaiaDR3,MaizApellaniz2022}
corresponds to roughly 1\,kpc but is too uncertain to anchor the
photospheric solid angle.

\begin{table*}[ht]
\caption{The four targets.\label{tab:sample}}
\centering
\begin{tabular}{lccccccc}
\hline\hline
Source & RA & Dec & $G$ & $d$ & $E(B-V)$ & Spectroscopy & Exposure \\
       & (deg) & (deg) & (mag) & (pc) & (mag) & & (s) \\
\hline
DESI~J000558.72$+$294103.8 & 1.49466  & $+$29.68439 & 19.37 & $604\pm98$  & 0.042 & DESI DR1 & co-add \\
DESI~J002253.23$+$134040.7 & 5.72177  & $+$13.67799 & 17.80 & $583\pm38$  & 0.068 & DESI DR1 & co-add \\
DESI~J074917.11$+$365427.9 & 117.32130 & $+$36.90776 & 19.17 & $946\pm135$ & 0.054 & DESI DR1 & co-add \\
LAMOST~J003553.36$+$433341.4 & 8.97235 & $+$43.56152 & 19.59 & $\sim$1000 & 0.058 & LAMOST DR11 & 5400 \\
\hline
\end{tabular}
\tablefoot{Distances are photogeometric distances based on \textit{Gaia} DR3
parallaxes \citep{GaiaDR3,BailerJones2021}; $E(B-V)$ values are from the
\citet{Schlafly2011} recalibration of the SFD dust map. For LAMOST~J0035$+$4333 the
zero-point-corrected parallax is $1.01\pm0.43$\,mas
\citep{MaizApellaniz2022}, i.e.,\ a distance of order 1\,kpc with
$\sim$40\% uncertainty; we therefore quote it for orientation only and
do not use it in the fit.}
\end{table*}

\subsection{ZTF photometry}
\label{sec:data_phot}

We retrieved archival light curves from ZTF DR23 \citep{Masci2019} within
$2''$ of each target, spanning 2018--2025 (baselines of 2500--2740\,d).
We retained only measurements with \texttt{catflags}$=0$ and magnitude
uncertainty $<0.30$\,mag.
The numbers of surviving $g/r/i$-band epochs are listed in
Table~\ref{tab:timing}; DESI~J0022$+$1340 is sparsely covered
($\sim$240 useful epochs), which limits the timing analysis of this
source (Sect.~\ref{sec:res_J0022}).

\subsection{Archival multiwavelength data}
\label{sec:data_mw}

We cross-matched the four targets against archival UV,
infrared, and X-ray catalogs: the Galaxy Evolution Explorer (GALEX),
the Two Micron All Sky Survey (2MASS), the AllWISE catalog of the Wide-field Infrared Survey Explorer (WISE), and the master tables
of the High Energy Astrophysics Science Archive Research Center
(HEASARC). The wider set of catalogs used for the spectral energy
distributions is introduced in Sect.~\ref{sec:sed}.
The results enter the per-source discussion of
Sect.~\ref{sec:results}; in brief:
DESI~J0022$+$1340 has a blue GALEX counterpart
($\mathrm{FUV} = 19.69 \pm 0.12$, $\mathrm{NUV} = 20.52 \pm 0.13$) and
was discovered as the optical transient MASTER~OT~J002253.29$+$134041.0
\citep{ATel6913};
DESI~J0749$+$3654 has a faint near-UV (NUV) counterpart
($22.11 \pm 0.36$) and is among the known CVs characterized in the
surveys of the eROSITA telescope (extended ROentgen Survey with an Imaging Telescope Array) on the Spektrum-Roentgen-Gamma
(SRG) mission \citep[][]{Schwope2024erosita};
LAMOST~J0035$+$4333 coincides (separation $3.6''$) with the slew-survey
X-ray source XMMSL3~J003553.6$+$433344
($F_{0.2-12\,\mathrm{keV}} = (1.8\pm0.9)\times10^{-12}$
$\mathrm{erg\,s^{-1}\,cm^{-2}}$) and has an AllWISE counterpart with a
mid-infrared excess ($W1-W2 = 0.69 \pm 0.10$); it has no 2MASS
counterpart, the nearest cataloged source lying $7.2''$ away;
DESI~J0005$+$2941 has no archival UV or X-ray coverage at its position
and was previously cataloged only as a WD candidate
\citep{GentileFusillo2019}.

\begin{table*}[ht]
\caption{ZTF photometry and timing results.\label{tab:timing}}
\centering
\begin{tabular}{lcccl}
\hline\hline
Source & $N_{g}/N_{r}/N_{i}$ & $\Delta m$ & $P_{\rm phot}$ & Notes \\
       &  & (mag) & (min) & \\
\hline
J0005$+$2941 & 1284/1614/259 & 2.5 & $105.349(3)$  & eclipse-like dip ($g,r$) \\
J0022$+$1340 & 77/126/37     & 2.0 & ---           & sampling-limited \\
J0749$+$3654 & 667/1927/134  & 1.5 & $40.2031(4)$  & double-humped; $P_{\rm orb}=80.406$ \\
J0035$+$4333 & 916/1206/187  & 2.5 & $143.765(5)$  & 1-d alias 159.7 not excluded \\
\hline
\end{tabular}
\tablefoot{Baselines are 2500--2740\,d for all four targets.
$P_{\rm phot}$ uncertainties (in parentheses, last digit) correspond to
the frequency resolution $1/T_{\rm baseline}$; the listed periods
exceed their 99\% bootstrap false-alarm levels in the bands given in
the notes ($g$ and $r$ unless stated otherwise).
$\Delta m$ is the full range of the rolling-median light curve,
reflecting accretion-state changes.}
\end{table*}

\section{Analysis}
\label{sec:methods}

\subsection{Joint continuum--cyclotron model}
\label{sec:joint}

The theory of cyclotron emission from the post-shock flows of magnetic
CVs is long established
\citep{1959westfold,1960SvPhU...3..115G,1962BB,1980chan,1981chanmugam,1982MW};
we did not re-derive it here.
Our forward model follows the standard isothermal, constant-$\Lambda$
prescription \citep[the uniform-temperature cyclotron grids
of][]{1985WM,wickramasinghe2000}, in which the emergent intensity is
\begin{equation}
I_\omega = I_\omega^{\rm RJ}\,\bigl[1-\exp(-\Psi_\omega \Lambda)\bigr],
\label{eq:cyc_intensity}
\end{equation}
where $I_\omega^{\rm RJ}$ is the Rayleigh--Jeans intensity at electron
temperature $kT$, $\Psi_\omega$ is the angle- and frequency-dependent
dimensionless cyclotron absorption coefficient for a mildly relativistic
thermal plasma, and $\Lambda = (\omega_p^2/\omega_c c)\,\ell$ is the
dimensionless size (optical depth) parameter of a column of thickness
$\ell$ \citep{1982MW,Vaeth1995,2020kolbina}.

The spectral shape is controlled by four parameters: the field strength
($B$), the viewing angle ($\theta$) between the line of sight and the field
(peak positions and widths), the electron temperature ($kT$; thermal
broadening), and the dimensionless optical depth parameter ($\Lambda$),
which sets how many of the low harmonics are optically thick.
The harmonics fall near
$\lambda_n \simeq (10\,710/n)\,(100\,\mathrm{MG}/B)$\,\AA{}
\citep[e.g.,][]{2022MNRAS.511...20K,2025A&A...696A.242V}, so the hump
spacing fixes $B$ largely independently of the other three parameters ---
the property that makes cyclotron spectroscopy a robust field diagnostic.

The observed spectrum is modeled as
\begin{equation}
\begin{split}
F_\lambda ={}& s_{\rm wd}\,K_\lambda(T_{\rm wd}, \log g)
    + s_{\rm spot}\,B_\lambda(T_{\rm spot}) \\
    &+ A_{\rm cyc}\,C_\lambda(B, kT, \theta, \Lambda),
\end{split}
\label{eq:model}
\end{equation}
where $K_\lambda$ is a pure-hydrogen WD (spectral type DA)
atmosphere model \citep{Koester2010}
interpolated in $(T_{\rm wd}, \log g)$, $B_\lambda$ is the hot-spot blackbody
representing the accretion-heated pole cap, and $C_\lambda$ is the
peak-normalized cyclotron component.
A common exploratory approach is to subtract a continuum (or a donor
template) fitted to presumed hump-free regions and then match the
residual to a cyclotron model -- by eye for survey-discovered systems
\citep[e.g.,][]{2025A&A...696A.242V} and by $\chi^2$ minimization for the
best-observed ones \citep[e.g.,][]{2022MNRAS.511...20K}.
Fitting the continuum and cyclotron components together instead avoids the
circular choice of those windows, propagates the continuum uncertainty
into the cyclotron parameters, and yields formal errors on $B$;
Sect.~\ref{sec:compare} quantifies the difference.

Before fitting, each spectrum is corrected for Galactic reddening
\citep[][with $R_V=3.1$]{CCM1989} using the $E(B-V)$ values of
Table~\ref{tab:sample}; strong emission lines (Balmer series through
H10, He\,I, He\,II~$\lambda$4686, Na\,I, and Ca\,II)
and telluric bands are masked rather than interpolated across, and the
masked spectrum is averaged into 25\,\AA{} bins weighted by the inverse
variance.
The fit covers 3950--9300\,\AA: below 3950\,\AA{} the crowded
high-order Balmer lines and the Balmer jump make the masking
incomplete and the survey flux calibration is least reliable.
Bin uncertainties combine the propagated pixel variance, the
intra-bin scatter, and a 3\% calibration floor.
No positivity constraint is imposed on the binned fluxes: after the
continuum is subtracted, bins scatter on both sides of the cyclotron
model within their errors, as expected for un-clipped data.
The model contains no donor component: none of the four spectra shows
TiO band heads or other late-type features. The donors are instead
constrained separately, from the archival mid-infrared photometry
(Sect.~\ref{sec:sed}); a possible low-level red-end contribution in
J0022$+$1340 is discussed in Sect.~\ref{sec:res_J0022}.
The fit statistic is the standard inverse-variance-weighted
\begin{equation}
\chi^2 = \sum_i \frac{\bigl[F_i - F_\lambda(\lambda_i)\bigr]^2}{\sigma_i^2}.
\label{eq:chi2}
\end{equation}

The nonlinear parameters $(T_{\rm wd}, \log g, T_{\rm spot}, B, kT,
\theta, \log\Lambda)$ are explored with differential evolution
\citep{StornPrice1997,Virtanen2020}; at every trial the three amplitudes
$(s_{\rm wd}, s_{\rm spot}, A_{\rm cyc})$ are solved exactly by bounded,
error-weighted linear least squares.
For the DESI sources, whose spectra are flux-calibrated, the WD solid
angle is bounded by the \textit{Gaia} distance:
$s_{\rm wd} = (R_{\rm wd}/d)^2$ with
$R_{\rm wd} \in [0.003, 0.030]\,R_\odot$ and $d$ within $\pm2\sigma$ of
the geometric distance.
This anchors the photospheric component without an external spectral energy distribution fit, but
does not tie $s_{\rm wd}$ to $\log g$ through a mass--radius relation
(Sect.~\ref{sec:wdprior}).
The WD photosphere and the hot-spot blackbody are smooth continua
of similar optical shape, so their temperatures and flux ratios are
mutually degenerate; we treated $T_{\rm wd}$, $T_{\rm spot}$, $s_{\rm wd}$,
and $s_{\rm spot}$ as nuisance parameters whose free-fit values we did not
interpret physically, and a smooth
component fitted with zero amplitude simply means the optical data do not
require it. For well-resolved humps within a fixed harmonic branch this
degeneracy stays in the continuum and does not reach the field: removing the
spot component or fixing its temperature changes the recovered $B$ by less
than the quoted precision for every target, and the benchmark recoveries
(Sect.~\ref{sec:benchmarks}) confirm it is unbiased. The continuum decomposition
therefore feeds the uncertainty on $kT$ and $\Lambda$ (and the derived
luminosities and areas of Sect.~\ref{sec:physics}), but not the field
within a fixed branch.
Where the humps are weak or blended, by contrast, the continuum treatment
does affect which harmonic branch is preferred (Sect.~\ref{sec:compare}),
and that branch choice is then the dominant field uncertainty.
The fitted electron temperature $kT$ ranges from a few tenths of a keV to
tens of keV across the sample and between branches because it, too,
trades against the continuum decomposition, so we read it as indicative
only. The low values of the adopted, cyclotron-dominated branches are
those expected when cyclotron radiation controls the post-shock cooling
(Sect.~\ref{sec:physics}), whereas the high $kT$ of some high-field
alternative branches is a symptom of the pseudo-continuum degeneracy
(Sect.~\ref{sec:systematics}) rather than a genuinely hotter shock; in
neither case does it affect the field, which is set by the hump spacing.

\paragraph{Harmonic ambiguity}
To expose discrete solution families, we computed the profiled curve
$\tilde\chi^2(B) = \min_{kT,\theta,\Lambda,\ldots}\,\chi^2$ on a grid of
$B$ from 12 to 96\,MG (step 2\,MG).
At every grid point all the other parameters, the continuum components
included, are re-optimized from several starting points.
We call this continuum-inclusive profile the full-refit profile, as
opposed to one computed over a fixed continuum.
Around the adopted minimum the profile is resampled at a 0.25\,MG
step.
We also computed the same profile at a series of fixed electron
temperatures, which separates the $B$ constraint from the $kT$
degeneracy and shows how the depth of the $\chi^2$ minimum responds to
the temperature; both families of curves are shown for every source in
Sect.~\ref{sec:results}.
We first collected every local minimum within $\Delta\chi^2 < 30$ (raw) of
the global one -- a deliberately wide net so that no plausible branch is
missed before rescaling -- and re-optimized each.
We then kept as an acceptable branch any that stayed within
$\Delta\tilde\chi^2 < 9$ ($3\sigma$) after the error rescaling described
below and shows a structured cyclotron component (peak-to-trough contrast
$>0.15$; Sect.~\ref{sec:systematics}).
The same thresholds are applied to the benchmarks and to all four
targets, with no per-object tuning. Varying them within reasonable limits
changes only which marginal solutions are listed, not the unique
J0005$+$2941 field or the main branch alternatives of the ambiguous
targets.
This makes the $(B, n)$ degeneracy, when present, visible in a
single diagnostic figure, rather than leaving the alternative harmonic
identification implicit.
For spectra with a single acceptable branch, we adopted the global minimum
as the field. When more than one branch satisfies the criteria, we quote
the lowest-$\chi^2$ branch as the formal global minimum and list the
competing acceptable branches; in those cases the branch spacing is part
of the result rather than a small statistical error.
Which branches are acceptable, and the precise position of the minimum,
depend on the fitted wavelength range and on the continuum and
viewing-geometry assumptions (Sect.~\ref{sec:systematics}). We fit the
science spectra over the full 3950--9300\,\AA{} band, restricting the
window to the cyclotron-dominated region only for the benchmarks, where
the band edges are dominated by continuum-subtraction artifacts
(Sect.~\ref{sec:benchmarks}); the contrast criterion above rejects
pseudo-continuum branches. When several branches are acceptable, the true
uncertainty on $B$ is how far apart those competing fields lie, which is
much larger than the formal error within any one branch.

\paragraph{Uncertainties}
Statistical uncertainties come from a Markov chain Monte Carlo (MCMC)
exploration \citep[\texttt{emcee};][]{ForemanMackey2013} of all ten
parameters of Eq.~(\ref{eq:model}): the seven nonlinear parameters and
the three component amplitudes.
The amplitudes, solved by linear least squares during the optimization
and the $\tilde\chi^2(B)$ scans, are here sampled as free parameters under
the same non-negativity bounds, so the intervals on $B$, $kT$, $\theta$,
and $\Lambda$ carry the full covariance between the continuum level and
the cyclotron shape.
The binned spectra scatter about the best model by more than their formal
errors (reduced $\chi^2 > 1$, as expected for survey data fitted with a
three-component model); we therefore widen every error bar by
$\sqrt{\chi^2_{\rm min}/{\rm dof}}$ before quoting intervals so that they
absorb this excess scatter.
As an independent check, the $\Delta\chi^2 < 1$ range of $B$ on the
rescaled profile agrees with the MCMC widths for every source.
Part of the excess is not random noise but model misfit: the isothermal,
single-region model cannot fully reproduce the hump shapes (e.g.,\ the
coherent residual in J0005$+$2941; Sect.~\ref{sec:res_J0005}). These
intervals are therefore a lower bound on the true uncertainty, conditional
on the emission model and the adopted harmonic branch
(Sect.~\ref{sec:systematics}). For the field in particular they are
subdominant: the realistic error on $B$ is set by the harmonic assignment
and the benchmark-calibrated systematic, not by the formal posterior
width.

\paragraph{Implementation}
The cyclotron intensity is evaluated in a form that stays accurate down
to low electron temperatures, a few tenths of a keV, where the usual
expressions become inaccurate; this matters for the cool, low-accretion
emission regions of polars such as DESI~J0005$+$2941.
The code and the data products needed to reproduce every fit are
available in the project repository\footnote{\url{https://github.com/wangnengdejiamao/cyclotron-spectrum-analyzer}}.

\subsection{Benchmark validation}
\label{sec:benchmarks}

\begin{figure*}[ht]
\centering
\includegraphics[width=0.9\linewidth]{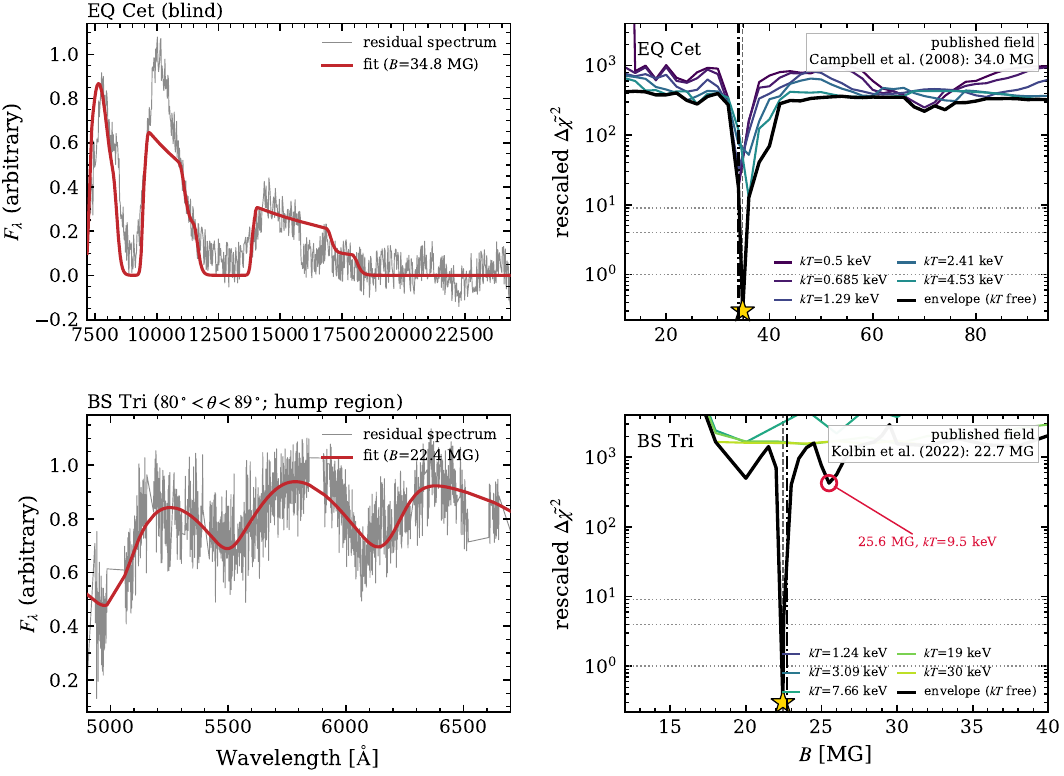}
\caption{Benchmark recovery on two polars with secure published fields.
\textit{Top row}: EQ~Cet, fitted blind to its phase-differenced infrared
residual spectrum, recovering $B=34.8$\,MG (published 34\,MG).
\textit{Bottom row}: BS~Tri, with the viewing angle restricted to
$80^\circ$--$89^\circ$ and the fit restricted to the cyclotron-hump
region (4900--6700\,\AA), recovering $B=22.4$\,MG
(published $22.7\pm0.4$\,MG).
\textit{Left column}: Phase-differenced residual spectrum (gray, full resolution)
over the fitted range, with the best-fit cyclotron model (red).
\textit{Right column}: $\tilde\chi^2(B)$ at fixed electron temperatures
(colored, see the legend in each panel) with the $kT$-free envelope
(black), drawn exactly as the source panels of Fig.~\ref{fig:B_profiles}. The dash-dotted lines mark the published field, labeled with its source,
and the stars the adopted solution at the envelope minimum.
The crimson circle in the BS~Tri panel marks the competing
$kT = 9.5$\,keV, $B = 25.6$\,MG solution of Sect.~\ref{sec:benchmarks},
which is $414$ above the adopted minimum after rescaling.
The plotted spectra, models, and $(B, kT)$ grids are included in the released data products (see Data availability).
\label{fig:bench}}
\end{figure*}

Having defined the model and the fitting, we tested the implementation and
the weighted-likelihood procedure on two benchmark polars with secure
published fields.
For EQ~Cet ($B \approx 34$\,MG) we fit the phase-differenced infrared
spectrum of \citet{2008campbellb}
--- the difference between the phases at which the accretion region is in
view and those at which it is hidden, not a phase average -- and recovered
$B = 34.8$\,MG, with a single acceptable minimum in the profiled
$\tilde\chi^2(B)$ curve (Fig.~\ref{fig:bench}); the only substantive
difference from their analysis is the well-known
$kT$--$\Lambda$ degeneracy.
For the eclipsing polar BS~Tri ($B = 22.7\pm0.4$\,MG,
$\theta\approx87^\circ$) we fit the phase-differenced residual spectrum of
\citet{2022MNRAS.511...20K} with a free viewing
angle, as they did; it converges to
$\theta\approx86^\circ$, in agreement with the eclipse geometry, so the
angle is recovered from the data rather than imposed. That agreement
presumes a magnetic axis roughly perpendicular to the orbital plane; if
the magnetic axis is inclined to the WD spin axis, or the spin
axis to the orbital angular momentum, $\theta$ need not track the orbital
inclination and can differ widely from it even in an eclipsing system.
Over the cyclotron-hump region (4900--6700\,\AA), where the
constant-$\Lambda$ model applies, the profiled $\tilde\chi^2(B)$ bottoms
at $B = 22.4$\,MG (Fig.~\ref{fig:bench}), matching the published field;
the $kT$--$B$ degeneracy also admits a higher-temperature
($kT = 9.5$\,keV) solution near $25.6$\,MG, but with a substantially
higher $\chi^2$ ($\chi^2/{\rm dof} = 2.19$ against $1.73$ on the same
1560 points, i.e.,\ $\Delta\chi^2 = 716$); a single blind fit
can land on it, whereas the profiled $\tilde\chi^2(B)$ selects the
published-field branch. It is marked in Fig.~\ref{fig:bench}.
The blue and red ends of this phase-differenced residual are dominated
by continuum-subtraction artifacts rather than cyclotron emission, and
including them raises the reduced $\chi^2$ from $\sim$1.7 to $\sim$4, so
the goodness of fit itself marks them as uninformative and sets the
fitted window.
Applied to the science targets, the same $\chi^2$-driven criterion
retains the full optical band, whose edges are not artifact-dominated
(Sect.~\ref{sec:systematics}).
As a further check on real survey data, we applied the identical
full-refit analysis to archival SDSS spectra of three more polars with
published cyclotron fields (Appendix~\ref{app:sdss}): MQ~Dra and PZ~Vir
are recovered within a few percent of their literature values
($\approx$59 and 63\,MG), while SDSS~J1344$+$2044 lands on a lower
harmonic branch (50.5\,MG against a published $56\pm2$\,MG), reproducing
the branch-ambiguity systematic that affects the science targets. The
field is therefore reliably recovered when the humps are well resolved
and over-determined, and is branch-limited otherwise.

\subsection{The white dwarf component}
\label{sec:wdprior}

In Eq.~(\ref{eq:model}) the photospheric temperature, surface gravity and
solid angle $s_{\rm wd} = (R_{\rm wd}/d)^2$ are free and mutually
independent, so the fit is not required to return a self-consistent star.
Table~\ref{tab:continuum} lists the values they take, with the radius that
$s_{\rm wd}$ implies at the \textit{Gaia} distance and the mass
$M_{\rm wd} = gR_{\rm wd}^2/G$ that follows. The combinations are
unphysical: DESI~J0005$+$2941 returns $T_{\rm wd} = 6000$\,K,
$\log g = 9.25$ and $R_{\rm wd} = 0.024\,R_\odot$, i.e.,\ $37\,M_\odot$,
and DESI~J0749$+$3654 $17\,M_\odot$, while in DESI~J0022$+$1340 the
photosphere runs to the lower bound of its solid angle and supplies 0.1\%
of the optical flux. This is the continuum degeneracy of
Sect.~\ref{sec:joint} in its most direct form: the data constrain the sum
of the two smooth components, not the division between them, so the free
decomposition is phenomenological and its WD parameters should
not be interpreted.

We therefore repeated every fit with the photosphere parametrized by mass
alone: $R_{\rm wd}$ from the \citet{Nauenberg1972} mass--radius relation,
$g = GM_{\rm wd}/R_{\rm wd}^2$, and $s_{\rm wd}$ from that radius and the
\textit{Gaia} distance, with $M_{\rm wd}\in[0.6,1.0]\,M_\odot$ and
$T_{\rm wd}\in[8000,20000]$\,K, representative ranges for CV WDs
around the $0.81\,M_\odot$ mean mass of \citet{Pala2022}; for
LAMOST~J0035$+$4333, whose flux scale is relative, only the temperature
range applies. These solutions are listed beside the free ones in
Table~\ref{tab:continuum} and shown in Appendix~\ref{app:wdprior}. The
photosphere now supplies 2--11\% of the optical flux of the three DESI
targets, as expected for a $\sim$10$^4$\,K WD at these distances,
at a cost of $\Delta\chi^2 = 5$, 34, 19 and 42, and the field is
unchanged: it shifts by less than $0.01$\,MG at the adopted branch of each
DESI target, by $0.4$\,MG at the alternative 77.7\,MG branch of
DESI~J0022$+$1340, and by $1.6$\,MG for LAMOST~J0035$+$4333, which is
movement along the shallow 51--54\,MG ridge of that spectrum
(Sect.~\ref{sec:res_J0035}). The constraint is not neutral about competing
branches, however: the 77.7\,MG branch of DESI~J0022$+$1340, essentially
tied with the adopted one in the free fit, falls to
$\Delta\tilde\chi^2 \approx 5$: both remain acceptable and we report both.
We quote the free fits throughout, since the benchmark validation, the
branch scan and the posterior sampling rest on them and the field is the
same either way, and read the constrained fits wherever the WD
and hot-spot temperatures or their flux ratio are at issue.

\subsection{Time-series analysis}
\label{sec:timing}

\begin{figure*}[ht]
\centering
\includegraphics[width=0.82\linewidth]{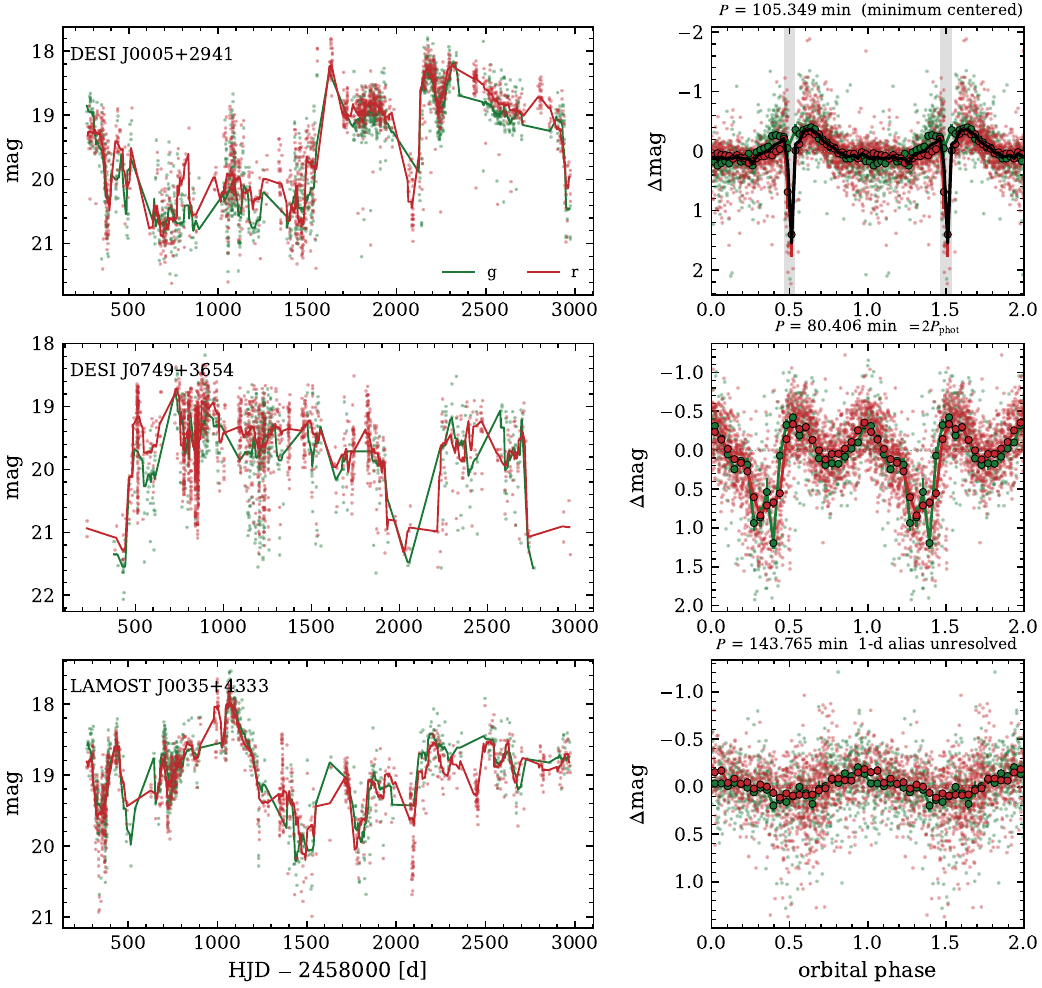}
\caption{ZTF light curves of the three targets with a significant period
($g$ in green, $r$ in red; the J0005$+$2941 folded panel also includes a
combined $g+r$ binned median in black).
\textit{Left}: Long-term light curves with same-color 40-day rolling
medians, showing the changes between bright and faint accretion states.
\textit{Right}: Detrended light curves folded at the adopted period
(J0749$+$3654 at the orbital period), with phase-binned median trend
lines. The J0005$+$2941 folded panel is phase-shifted to center its
narrow eclipse-like minimum at phase~0.5. J0749$+$3654 shows the double
cyclotron-beaming hump per orbit, and J0035$+$4333 a quasi-sinusoidal
modulation.
DESI~J0022$+$1340, which has no significant period, is not shown.
\label{fig:lightcurves}}
\end{figure*}

Polars vary on the orbital period through cyclotron beaming and
self-eclipses of the accretion region, and additionally switch between
high and low accretion states on timescales of months to years.
Untreated state changes corrupt period searches, so we first removed
long-term trends by subtracting a 40-day rolling median from each band
(a window far longer than the minute-to-hour orbital periods, so
state changes are removed while the orbital modulation is preserved),
and inspected each light curve for low states (intervals fainter than
0.8\,mag below the bright-state median).
The detrended $g$, $r$, and (where available) $i$ data were searched with
the generalized Lomb--Scargle periodogram
\citep{Lomb1976,Scargle1982,VanderPlas2018} over 2--40\,cycles\,d$^{-1}$
(36\,min--12\,hr), per band and for the variance-normalized combination
of bands.
False-alarm levels are estimated by bootstrap permutation of the
magnitudes.
For every candidate frequency we tabulated the power of its $\pm1$ and
$\pm2$\,d$^{-1}$ aliases, its harmonic and subharmonic, and we required
consistency between independent bands before a period is considered
secure.
Folded light curves at the adopted period and at its strongest alias are
inspected for each source.
All four targets show large accretion-state changes between bright and faint states, with
full amplitudes of 1.5--2.5\,mag on month-to-year timescales and without
the fast-rise, slow-decline morphology of dwarf-nova outbursts;
Fig.~\ref{fig:lightcurves} shows the long-term and folded light curves of
the three with a significant period.

\begin{table*}[ht]
\caption{Continuum parameters of the adopted fits and of the physically constrained refit.\label{tab:continuum}}
\centering
\begin{tabular}{llccccccc}
\hline\hline
Source & Fit & $T_{\rm wd}$ & $\log g$ & $R_{\rm wd}$ & $M_{\rm wd}$ & $T_{\rm spot}$ & WD/spot/cyc & $B$ \\
 & & (K) & (cgs) & ($R_\odot$) & ($M_\odot$) & (K) & (percent) & (MG) \\
\hline
DESI~J0005$+$2941 & free & 6000 & 9.25 & 0.0240 & 37 & 21275 & 11/72/17 & 56.77 \\
 & constrained & 8000 & 8.21 & 0.0101 & 0.60 & 17916 & 11/73/16 & 56.77 \\
\hline
DESI~J0022$+$1340 & free & 6000 & 7.00 & 0.0027 & 0.003 & 22671 & 0/0/100 & 64.35 \\
 & constrained & 8000 & 8.88 & 0.0060 & 1.00 & 30379 & 2/0/98 & 64.35 \\
\hline
DESI~J0749$+$3654 & free & 10337 & 8.50 & 0.0386 & 17 & 10342 & 61/35/3 & 29.26 \\
 & constrained & 10832 & 8.21 & 0.0101 & 0.60 & 10832 & 9/87/3 & 29.26 \\
\hline
LAMOST~J0035$+$4333 & free & 6000 & 7.00 & \ldots & \ldots & 99204 & 81/6/13 & 51.13 \\
 & constrained & 20000 & \ldots & \ldots & \ldots & 50000 & 0/34/66 & 52.69 \\

\hline
\end{tabular}
\tablefoot{The ``free'' rows are the adopted solutions of Table~\ref{tab:results}, in which $T_{\rm wd}$, $\log g$ and the solid angle $s_{\rm wd}=(R_{\rm wd}/d)^2$ vary independently. $R_{\rm wd}$ is then the radius implied by $s_{\rm wd}$ at the \textit{Gaia} distance, and $M_{\rm wd}=gR_{\rm wd}^2/G$ the mass implied by that radius together with the fitted gravity. The free-fit masses of the three DESI targets lie outside the WD range, so the free fit does not return a physical WD (Sect.~\ref{sec:wdprior}). The ``constrained'' rows impose the \citet{Nauenberg1972} mass--radius relation with $M_{\rm wd}\in[0.6,1.0]\,M_\odot$, $T_{\rm wd}\in[8000,20000]$\,K and the \textit{Gaia} distance, so $\log g$, $R_{\rm wd}$ and $s_{\rm wd}$ all follow from $M_{\rm wd}$. LAMOST~J0035$+$4333 has a relative flux scale, so only the temperature priors apply to it; we thus cannot infer the gravity, radius, or mass. Its free-fit hot-spot temperature is a legacy solution that predates the $5\times10^{4}$\,K cap now enforced by the released code, and is superseded by the constrained row. The penultimate column gives the band-integrated flux fraction of the three components over 3950--9300\,\AA, and the last column the recovered field.}
\end{table*}

\section{Results}
\label{sec:results}

\begin{table*}[ht]
\caption{Joint-fit cyclotron parameters and ZTF timing results.
\label{tab:results}}
\centering
\begin{tabular}{llccccc}
\hline\hline
Source & Branch & $B$ & $kT$ & $\theta$ & $\log_{10}\Lambda$ & Period \\
       &        & (MG) & (keV) & (deg) & & (min) \\
\hline
DESI~J0005$+$2941 & unique & $\BJaaa\pm\BJaaaErr$ & 0.51 & 55 & 6.0 & $\PJaaa$ \\
DESI~J0022$+$1340 & global min. & $64.4\pm1.7$ & 4.2 & 47 & 4.5 & --- \\
                  & alt.\ branch & $77.7\pm2.0$ & 25.4 & 89 & 1.4 & --- \\
DESI~J0749$+$3654 & global min. & $29.3\pm0.8$ & 1.9 & 70 & 5.8 & $80.4062$ \\
                  & alt.\ branch & $96.4\pm2.7$ & 11.8 & 47 & 2.3 & $80.4062$ \\
LAMOST~J0035$+$4333 & single & $51.1\pm1.4$ & 2.6 & 63 & 4.2 & $\PJddd$ \\
\hline
\end{tabular}
\tablefoot{Fields are from the full-refit branch scan (continuum and
cyclotron re-optimized at every trial $B$). DESI~J0005$+$2941 has a
single robust solution and LAMOST~J0035$+$4333 a single but broad
minimum; DESI~J0022$+$1340 and DESI~J0749$+$3654 admit two distinct
acceptable harmonic branches (``global min.'' is the lowest-$\chi^2$
branch within the adopted single-region model; ``alt.'' is the competing
acceptable branch).
For DESI~J0749$+$3654 the listed period is the orbital period inferred
from the double-humped ZTF modulation. For LAMOST~J0035$+$4333 the table
lists the adopted 143.8\,min period; the 159.7\,min 1-day alias remains
possible and is discussed in Table~\ref{tab:timing} and Sect.~\ref{sec:res_J0035}.
Its profile has a shallow 51--54\,MG ridge within the same harmonic
assignment; the table gives the deeper minimum and its local uncertainty,
which does not extend to the 54\,MG sub-minimum.
The field error combines the formal statistical precision with a $2.6\%$
benchmark systematic floor, which dominates it (Sect.~\ref{sec:systematics});
for the two branch-ambiguous sources the branch spacing, far larger than
these errors, is the real field uncertainty.
$kT$, $\theta$, and $\log_{10}\Lambda$ are indicative values of the partly
degenerate continuum--cyclotron decomposition; the field within a fixed
branch is the robust quantity. The 45.3\,MG comparison branch of DESI~J0749$+$3654 is shown
only in Appendix~\ref{app:branches}. Periods are from
Table~\ref{tab:timing}.}
\end{table*}

\subsection{DESI J000558.72+294103.8}
\label{sec:res_J0005}

\paragraph{Spectroscopy}
The joint fit (Fig.~\ref{fig:joint_fits}) decomposes the DESI spectrum
into a cool WD photosphere, a hot-spot component, and a cyclotron
component with well-separated humps at $\approx$4700, 6200--6700, and
8900\,\AA, which correspond to cyclotron harmonics $n = 4$, 3, and 2
of a 56.8\,MG field (rest harmonic wavelengths 4717, 6290, and
9434\,\AA).
The profiled $\tilde\chi^2(B)$ curve (Fig.~\ref{fig:B_profiles}) has a
single acceptable minimum: alternative harmonic identifications are
excluded with $\Delta\chi^2 > 900$ (raw; $>170$ after error rescaling),
so the harmonic ambiguity that affects single-hump systems is absent
here.
The joint posterior gives $B = \BJaaa \pm \BJaaaErr$\,MG,
$kT = \TJaaa$\,keV, $\theta = \thJaaa$\,deg, and
$\log_{10}\Lambda = \lamJaaa$
(reduced $\chi^2 = \chiredJaaa$; intervals rescaled by
$\sqrt{\chi^2_{\rm red}}$).
The quoted field error is dominated by the $2.6\%$ systematic floor
calibrated from the benchmark recoveries
(Sect.~\ref{sec:systematics}); the formal statistical precision from the
rescaled posterior is much smaller ($\lesssim0.1$\,MG), reflecting the
three well-separated harmonics.
The fit leaves a coherent residual between the humps near 5500\,\AA:
the model there exceeds the data by about $1.3$ times the bin
uncertainty, with the same sign in every bin across that interval, so
the deviation is systematic rather than random
(Fig.~\ref{fig:joint_fits}). The isothermal model cannot fully reproduce
the hump-to-valley contrast.
Restricting the fit to the saturated regime ($\log\Lambda > 7.5$)
raises $\chi^2$ by $\approx 190$ and moves $B$ by only 0.3\,MG.
A two-step continuum-subtraction fit gives a similar field
($58.1$\,MG), while $kT$ and $\Lambda$ shift between branches and should
be read as model-dependent.
The low electron temperature is typical of optical cyclotron fits and
reflects the cyclotron-dominated cooling regime discussed in
Sect.~\ref{sec:physics}.
Before our DESI catalog the source had been listed only as a
WD candidate \citep{GentileFusillo2019}; \citet{2026hou} then
classified it as a polar and reported $B\approx45$\,MG from the uniform
two-step pipeline applied across that catalog. That value corresponds
to a different harmonic assignment, which is not selected by the present
profiled scan ($\Delta\chi^2>900$); the joint analysis here gives a
unique, well-resolved minimum at $\BJaaa$\,MG.
Its position lacks GALEX and pointed X-ray coverage, and it is the one
target whose WISE flux is accounted for by the donor expected at its
period (Sect.~\ref{sec:sed}).

\paragraph{Timing}
The ZTF light curve (3157 useful $g/r/i$ epochs over 2705\,d;
Fig.~\ref{fig:lightcurves}) shows the irregular high- and low-state
transitions characteristic of polars.
After detrending, all three bands independently select the same
frequency, $f = 13.6689$\,d$^{-1}$, i.e.,
$P = 105.349 \pm 0.003$\,min (refining the 1.76\,hr period of
\citealp{2026hou}), with bootstrap false-alarm probabilities
$\ll 10^{-3}$; the $\pm1$\,d$^{-1}$ aliases are weaker by a factor
$\gtrsim 1.6$ in power, and folding at the strongest alias produces no
coherent structure (Fig.~\ref{fig:lightcurves}).
The folded light curve displays a deep, narrow, eclipse-like feature.
In 0.02-wide phase bins its full width at half depth is 0.02--0.04 in
phase (2--4\,min); the depth is at least 1.4\,mag in $g$ and $r$
(a lower limit, because many in-eclipse epochs fall below the ZTF
detection threshold and are censored from the light curve) and only
$\sim$0.8\,mag in $i$, where the donor contributes residual flux.
A 0.3\,mag brightening hump follows the eclipse.
Several properties favor a geometric eclipse over a cyclotron
self-eclipse or beaming minimum: the feature is far narrower (2--4\,min)
than the broad, quasi-sinusoidal beaming minima of the other targets,
its depth is nearly achromatic between $g$ and $r$, and the shallower,
nonzero $i$-band minimum is consistent with the residual donor light
expected when the cool secondary occults the compact emission region.
If the feature is indeed a geometric eclipse, its width constrains the
inclination: for a Roche-lobe-filling donor on the \citet{knigge2011}
sequence at 105\,min ($M_2 \approx 0.11\,M_\odot$,
$R_2/a \approx 0.23$) and a WD mass of 0.6--0.8\,$M_\odot$, an
eclipse of the compact light source lasting 0.02--0.04 in phase suggests
$i \approx 77$--$80^\circ$ for this simple geometry; at $i = 90^\circ$ the
maximum eclipse width would be 0.07 in phase, comfortably above the
measured value.
The depth, narrowness, color dependence, and stability of the feature
over the 7\,yr baseline identify $P = 105.349$\,min as the orbital
period and make the system a strong eclipsing-polar candidate below the
period gap. The ZTF sampling does not resolve the eclipse profile, so we
kept the eclipse as a candidate; high-speed photometry that resolves its
ingress and egress would confirm it and tighten the inclination.
Eclipsing polars are rare -- only a few dozen are known, and
there are very few below the gap \citep{2025schwope} -- so finding one in a
survey-selected sample is notable.

\subsection{DESI J002253.23+134040.7}
\label{sec:res_J0022}

\paragraph{Spectroscopy}
The DESI spectrum of J0022$+$1340 is unusual: the joint fit attributes
essentially all of the optical flux to the cyclotron component, with
the best-fit model flux of the WD photosphere and hot-spot components
consistent with zero, and three broad humps near 4100, 5300, and
7400\,\AA{} dominating the spectrum (Fig.~\ref{fig:joint_fits}).
The profiled $\tilde\chi^2(B)$ curve has its formal global minimum at
$B = \BJbbb \pm \BJbbbErr$\,MG ($kT = \TJbbb$\,keV,
$\theta = \thJbbb$\,deg, $\log_{10}\Lambda = \lamJbbb$;
reduced $\chi^2 = \chiredJbbb$), where the humps are harmonics
$n = 4$, 3, and 2 with peaks displaced blueward of the rest harmonic
wavelengths by the thermal motion of the 4\,keV plasma.
A second, nearly degenerate branch at $77.7$\,MG is essentially tied with
this minimum ($\Delta\chi^2 \approx 0.07$ after rescaling;
Table~\ref{tab:results}, Fig.~\ref{fig:B_profiles}), so we report both
branches and refer to the 64.4\,MG branch only as the formal global
minimum.
Our DESI catalog \citep{2026hou} measured $B = 73.05$\,MG for this
source from the two-step fit, near the upper (77.7\,MG) branch -- a
direct example of how the continuum treatment selects between the
harmonic branches that the joint fit instead keeps open.
Two archival facts bear on its nature.
The system was discovered as the optical transient
MASTER~OT~J002253.29$+$134041.0, caught at 17.1\,mag against an
18--20.5\,mag quiescent range \citep{ATel6913}; such brightenings are
compatible either with accretion-state transitions of a polar or with
disk outbursts of an intermediate polar, and the sparse ZTF coverage
(below) does not distinguish the two morphologies.
GALEX also records a blue UV counterpart
($\mathrm{FUV}-\mathrm{NUV} = -0.8$), lying an order of magnitude above
the extrapolated optical model (Sect.~\ref{sec:sed}). A hot photospheric
or accretion-heated component is therefore present even though the
optical fit gives it a negligible flux: the WD is not absent;
it is outshone by cyclotron emission in the optical. Since the GALEX
epochs predate the spectrum by more than a decade and the system varies
by $\sim$2\,mag, this shows only that a hot component exists, not what
its temperature is.
Three caveats remain. First, the $B$ value rests entirely on the hump
positions. Second, the spectrum rises mildly redward of 8000\,\AA{}, where
a low-level donor contribution cannot be excluded, although the
mid-infrared limits it to M5\,V or later (Sect.~\ref{sec:sed}) and the
two bluer humps drive the field. Third, the magnetic subtype (polar
vs. intermediate polar) remains open.
Phase-resolved spectroscopy, polarimetry, or an X-ray observation of
this system would be particularly valuable.

\paragraph{Timing}
Only $\sim$240 ZTF epochs survive the quality cuts, distributed over
2500\,d with large seasonal gaps; the long-term light curve varies by
$\sim$2\,mag (Fig.~\ref{fig:lightcurves}), consistent with
accretion-state changes.
The Lomb--Scargle analysis of the detrended data does not yield a
significant period: the three bands select mutually inconsistent peaks
(zg: 718\,min; zr: 153\,min; zi: 46\,min), none of which exceeds its
99\% bootstrap false-alarm level convincingly, and the strongest
combined-band peaks fall at 2.01, 3.01, and 4.01\,d$^{-1}$, exact
harmonics of the 1-day sampling pattern.
The present photometry does not support a period determination for
this source.
An orbital period for DESI~J0022$+$1340 will require dedicated
time-series photometry.

\subsection{DESI J074917.11+365427.9}
\label{sec:res_J0749}

This system is also known as SDSS~J074917.11$+$365428.0 (MGAB--V3545), classified
as a polar by \citet{2023inighta}, detected in the SRG/eROSITA surveys
\citep{Schwope2024erosita}, and accompanied by a faint GALEX NUV
counterpart ($22.11\pm0.36$).
\citet{2023inighta} measured an orbital period of
$0.0558\pm0.0001$\,d ($80.35\pm0.14$\,min).
No reliable magnetic-field measurement has been published: our DESI
catalog \citep{2026hou} likewise found the cyclotron features
--- confined to the red end of the spectrum -- too weak to constrain
the field.

\paragraph{Spectroscopy}
The DESI spectrum of J0749$+$3654 does not uniquely determine the field.
A full-refit scan over $B$ finds statistically competitive
single-component solutions near 30 and 96\,MG; the $n=5$--3 assignment
near 45\,MG has a higher $\chi^2$ once the continuum is re-optimized freely
(Fig.~\ref{fig:B_profiles}; Appendix~\ref{app:branches}).
The lower-field branch matches the red residual structures near 7400 and
8900\,\AA{} but under-fits the blue, while the high-field branch tends
toward a broad pseudo-continuum, so none is decisive.
We therefore report J0749$+$3654 as branch-ambiguous and do not adopt a
single spectroscopic magnetic field from the DESI co-add alone.

This ambiguity is consistent with the light-curve morphology.
The folded ZTF data show two unequal cyclotron-beaming humps per orbit,
suggesting either a nonuniform accretion region or more than one
emitting region. A two-region or phase-resolved model may be needed
before a single field can be interpreted. Until such data exist, the
secure results for J0749$+$3654 are its magnetic-CV
classification and its precise orbital period, while its optical
cyclotron field should be quoted as branch ambiguous.

\paragraph{Timing}
The detrended ZTF data show a highly significant
signal at $f = 35.8183$\,d$^{-1}$, i.e.,\
$P_{\rm phot} = 40.2031 \pm 0.0004$\,min, in $g$ and $r$ independently;
twice this value, $2P_{\rm phot} = 80.4062 \pm 0.0008$\,min, agrees
with the published $80.35\pm0.14$\,min within its uncertainty and
sharpens it considerably.
Folded at $2P_{\rm phot}$ (Fig.~\ref{fig:lightcurves}), the light curve
shows two bright humps of unequal shape per orbit (phases
$\sim$0.3--0.45 and $\sim$0.7--0.85) separated by a deep
($\sim$1\,mag) minimum at phase 0.05--0.25.
This double-humped morphology explains why the periodogram selects half
the orbital period. It is naturally produced by cyclotron beaming from
the accretion region(s): emission maxima occur when the field lines at the
emitting pole(s) are viewed perpendicularly, twice per orbit for
favorable geometries, with the deep minimum corresponding to the
phase at which the dominant region points toward (or is hidden from)
the observer.
The 80.41\,min period places the system very close to the CV period
minimum, among the shortest-period polars known
\citep{2025schwope}.

\subsection{LAMOST J003553.36+433341.4}
\label{sec:res_J0035}

\begin{figure*}[ht]
\centering
\includegraphics[width=0.49\linewidth]{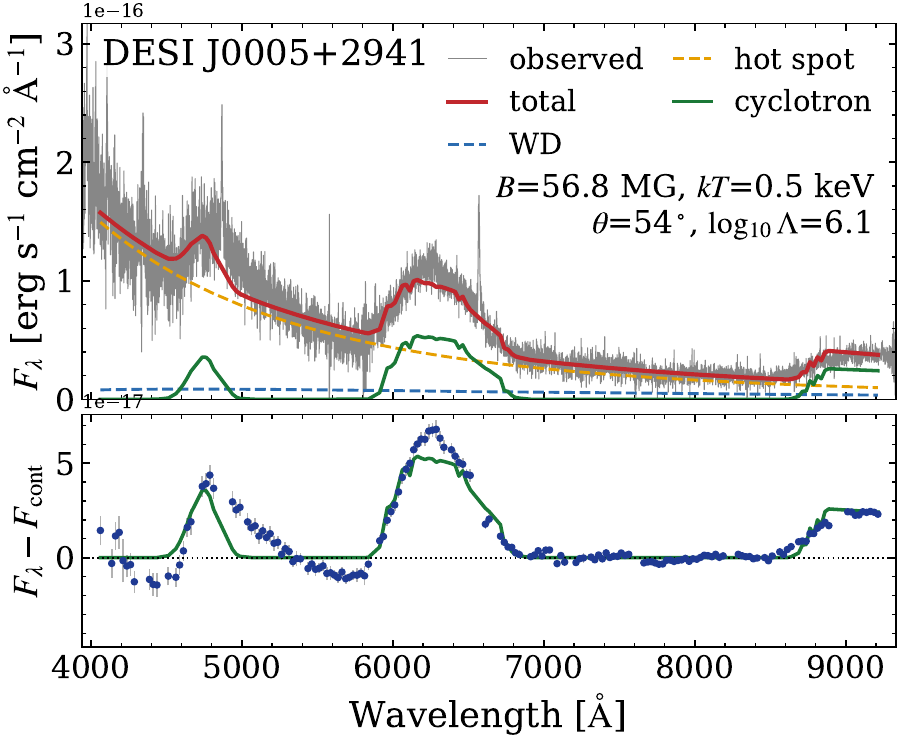}
\includegraphics[width=0.49\linewidth]{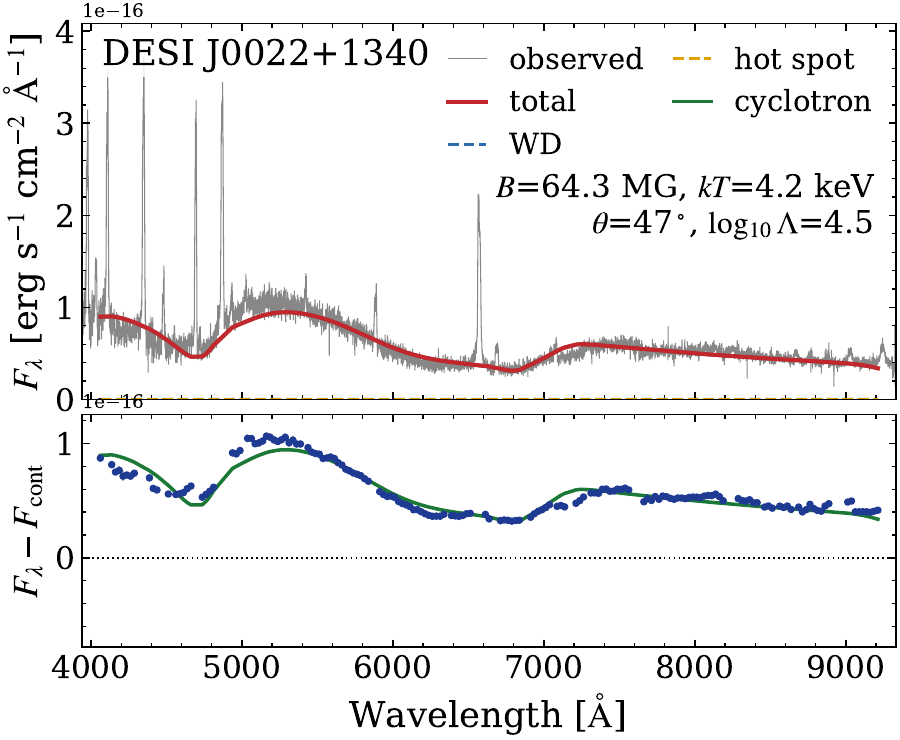}\\
\includegraphics[width=0.49\linewidth]{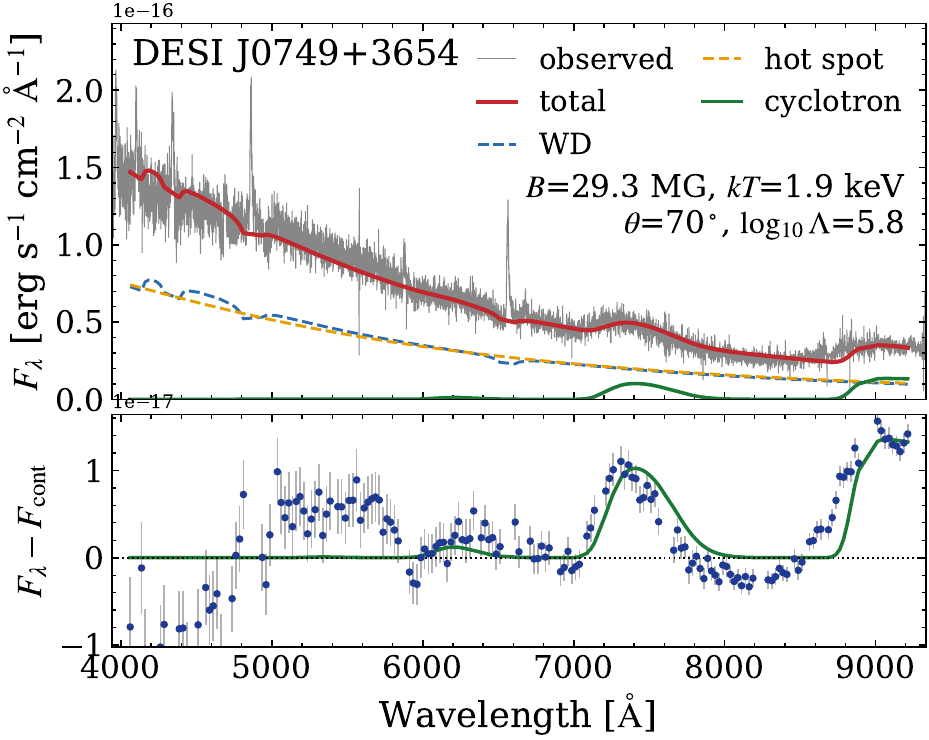}
\includegraphics[width=0.49\linewidth]{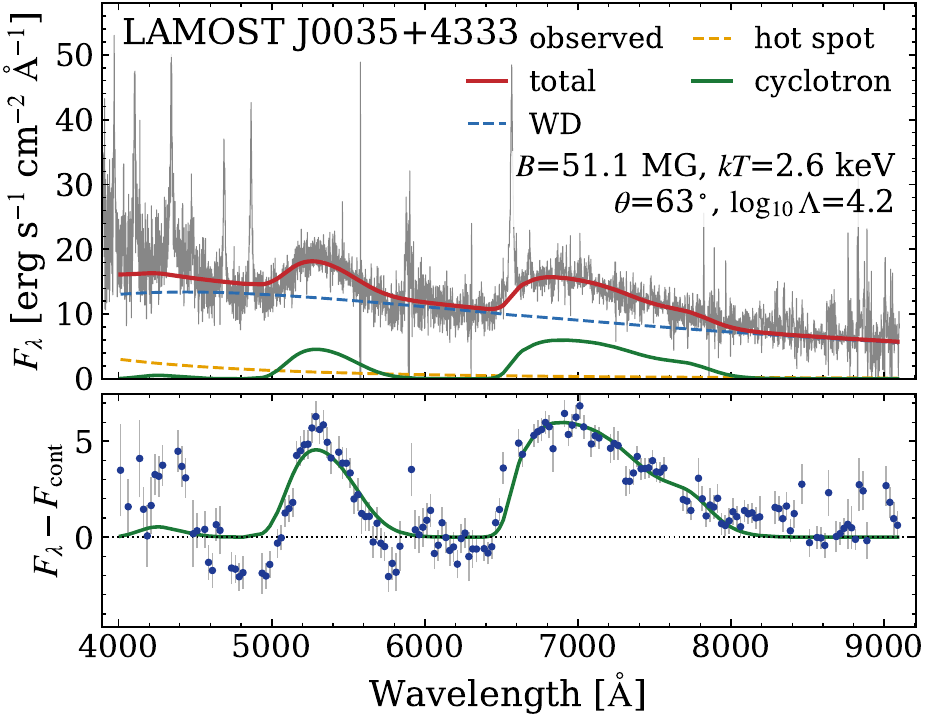}
\caption{Joint continuum--cyclotron fits for DESI~J0005$+$2941
(\textit{top left}), DESI~J0022$+$1340 (\textit{top right}), DESI~J0749$+$3654 (\textit{bottom
left}), and LAMOST~J0035$+$4333 (\textit{bottom right}).
In each panel, the upper plot shows the observed, de-reddened
spectrum (gray; emission lines are visible but masked in the fit)
with the total model (red), the WD photosphere (dashed blue),
the accretion-spot blackbody (dashed orange), and the cyclotron
component (green), all evaluated on the 25\,\AA{} fitting grid. The
lower plots show the binned data minus the continuum components,
with error bars, compared with the cyclotron component alone.
No continuum window selection or residual clipping is involved; bins
scattering below zero do so within their
uncertainties.
The fitted spectra and model components are included in the released data products (see Data availability).
Annotated parameters are the $\chi^2$-minimum best-fit values; their last digit can differ from the
posterior medians quoted in Table~\ref{tab:results}.
\label{fig:joint_fits}}
\end{figure*}

\begin{figure*}[ht]
\centering
\includegraphics[width=0.98\textwidth]{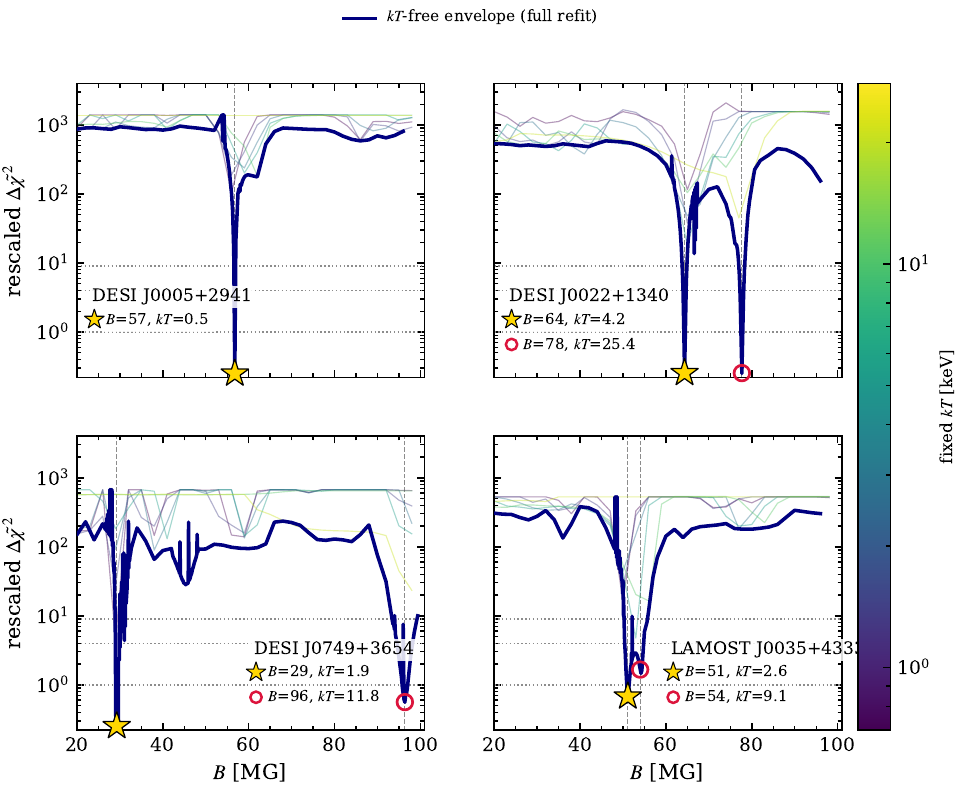}
\caption{Full-refit profiled $\chi^2(B)$ curves for the four science
targets --  DESI~J0005$+$2941
(\textit{top left}), DESI~J0022$+$1340 (\textit{top right}), DESI~J0749$+$3654 (\textit{bottom
left}), and LAMOST~J0035$+$4333 (\textit{bottom right}) -- in the same style as the EQ~Cet and BS~Tri benchmark
(Fig.~\ref{fig:bench}). Thin colored lines are profiles at fixed
electron temperatures, $kT$ (see the color bar). The thick navy curve is the
$kT$-free envelope, in which the WD, hot-spot blackbody, and all
cyclotron shape parameters are re-optimized and the amplitudes re-solved
at every $B$. All curves are rescaled by $\chi^2_{\rm min}/{\rm dof}$.
Filled stars mark the unique or formal lowest-$\chi^2$ branches; open circles mark the
second-best acceptable branches where a distinct alternative exists, and
the labels give each branch's $B$ and $kT$.
For these survey spectra, the fixed-$kT$ curves are mostly flat away from
the envelope minima -- the field is set by the hump spacing, not by the
assumed temperature -- unlike the strong-hump benchmarks.
DESI~J0005$+$2941 has a single robust minimum. DESI~J0022$+$1340 and
DESI~J0749$+$3654 show genuinely competing harmonic assignments (64 vs.\ 78\,MG
and 29 vs.\ 96\,MG, respectively), so their field uncertainty is dominated by the branch
choice rather than the formal posterior width (Appendix~\ref{app:branches}).
The two sub-minima of LAMOST~J0035$+$4333 (51 and 54\,MG) are instead the
same harmonic at the two ends of its $kT$--$B$ degeneracy, 3\,MG apart;
the quoted error is that of the deeper minimum and does not span both.
\label{fig:B_profiles}}
\end{figure*}

\paragraph{Spectroscopy}
The LAMOST spectra are calibrated only in relative flux and have the
lowest signal-to-noise ratio of the sample ($\mathrm{S/N}\sim4$--7 per
pixel), so the continuum decomposition is less secure than for the DESI
sources.
The joint fit to the 2014 spectrum nevertheless reproduces the
dominant structures near 4300, 5300, and 6900\,\AA{} (harmonics
$n = 5$--3 of a 51\,MG field) with $B = \BJddd\pm\BJdddErr$\,MG
(reduced $\chi^2 = \chiredJddd$; Fig.~\ref{fig:joint_fits}).
The associated $n=2$ harmonic would lie near 1.05\,$\mu$m, beyond the
fitted optical window, so it is constrained only indirectly by the
red-end continuum curvature.
Unlike the genuinely different harmonic branches of DESI~J0022$+$1340 and
DESI~J0749$+$3654, the profiled curve of this low-S/N spectrum shows two
close, nearly equal-$\chi^2$ sub-minima at $B \approx 51$ and $54$\,MG
(Fig.~\ref{fig:B_profiles}). These are not competing harmonic
assignments: they use the same harmonic numbers. Instead, they are two
points on a shallow $kT$--$B$ ridge, with $kT \approx 2.6$ and $9$\,keV
at the two minima and $\Delta\tilde\chi^2 \approx 2$ between them. We
adopted the deeper $51$\,MG minimum and quote its width, with the benchmark
systematic floor added in quadrature. That error does not reach the
$54$\,MG sub-minimum $2.9$\,MG away: along this ridge the field is
constrained only at the $\sim$3\,MG level.
Because the flux scale is relative and no distance prior can be
applied, we list $B = \BJddd\pm\BJdddErr$\,MG as indicative in
Table~\ref{tab:results}; a flux-calibrated spectrum of moderate S/N
would convert it into a measurement.
The thermodynamic parameters of this source should not be considered
constrained.

The archival record independently supports ongoing magnetic accretion:
the source coincides with the slew-survey X-ray detection
XMMSL3~J003553.6$+$433344
\citep[$F_{0.2-12\,\mathrm{keV}} = (1.8\pm0.9)\times10^{-12}$
$\mathrm{erg\,s^{-1}\,cm^{-2}}$;][]{Saxton2008}, which at the
$\sim$1\,kpc parallax distance corresponds to
$L_{\rm X} \sim 2\times10^{32}\,(d/1\,{\rm kpc})^{2}$
$\mathrm{erg\,s^{-1}}$, a typical high-state polar luminosity.
Its WISE counterpart is the reddest of the sample
($W1-W2 = 0.68\pm0.05$) and lies a factor of $\sim$3 above the donor
expected at its period. No M or early-L dwarf reaches $W1-W2 > 0.3$
\citep{PecautMamajek2013}, so the excess cannot be explained by the donor
alone and is likely accretion related, with cyclotron emission entering
the mid-infrared the natural candidate (Sect.~\ref{sec:sed}).

\paragraph{Timing}
The ZTF light curve (2309 epochs, 2700\,d) shows quasi-cyclic state
changes of up to 2.5\,mag (Fig.~\ref{fig:lightcurves}).
The detrended data yield a clear periodicity, but with a genuine
1-day-alias ambiguity. The $g$ band and the combined data prefer
$f = 10.0163$\,d$^{-1}$ ($P = 143.765 \pm 0.005$\,min). The $r$ band
alone prefers the $f-1$ alias at 159.7\,min, and the two combined peaks
differ by only $\sim$2\% in power.
Folded at 143.77\,min the light curve is a smooth, quasi-sinusoidal
modulation of $\sim$0.25\,mag amplitude in all bands, consistent with
cyclotron beaming from an accretion region that remains in view for most
of the orbit.
We adopted $P = 143.8$\,min as the most probable period, though the
159.7\,min alias cannot yet be ruled out. Either value places the system
within the 2--3\,hr period gap, which polars populate more readily than
nonmagnetic CVs \citep{2021schreiber,2024schreiber}.

\subsection{Emission-line diagnostics}
\label{sec:emission}

The cyclotron fit masks the emission lines, but they give an independent
test of the magnetic, accreting nature of the systems.
We measured the equivalent widths (EWs) and fluxes of the Balmer,
neutral-helium, and high-excitation He\,II~$\lambda$4686 lines by
local-continuum integration (Table~\ref{tab:emission},
Fig.~\ref{fig:emission}).
\citet{2026hou} already noted the emission-line activity of the three
DESI targets -- strongest in J0022$+$1340, whose Balmer, Paschen, and He
lines show $\sim$200\,km\,s$^{-1}$ velocity offsets between hydrogen and
helium -- which we quantify here.

\paragraph{High-excitation helium}
He\,II~$\lambda$4686, produced when the hard radiation of the
accretion column photoionizes the infalling gas, is the classical
optical signature of a magnetic WD \citep{Silber1992}.
It is strong in two spectra:
DESI~J0022$+$1340 has
EW(He\,II)$\,\approx\,$EW(H$\beta$) (ratio $1.0$, at
EW(H$\beta$)$\,=\,55$\,\AA), and LAMOST~J0035$+$4333 has a ratio of
$0.76$ at EW(H$\beta$)$\,=\,33$\,\AA.
Both satisfy the \citet{Silber1992} magnetic-activity criterion
(EW(H$\beta$)$\,>\,20$\,\AA{} and He\,II/H$\beta\,>\,0.4$),
independently confirming the magnetic classification implied by the
cyclotron humps.
In DESI~J0005$+$2941 the line is weak, and in DESI~J0749$+$3654 it is
marginal at best (ratios $0.25$ and $0.07$); since the He\,II strength tracks the
instantaneous accretion state and the hardness of the ionizing field, a
weak line at a single epoch does not argue against a magnetic nature.

\paragraph{Balmer decrement}
All four systems show a flat-to-inverted Balmer decrement:
H$\alpha$/H$\beta$ ranges from $0.7$ to $1.4$ and
H$\gamma$/H$\beta\,\approx\,0.8$--$1.0$.
Such flat decrements are commonly seen in accreting magnetic CVs and
indicate dense line-forming gas in the accretion flow rather than a
simple optically thin nebular spectrum
\citep{Williams1980,Echevarria1988,1995warner,wickramasinghe2000}.
The cyclotron humps form a curved pseudo-continuum beneath the lines, so
the absolute H$\alpha$/H$\beta$ values carry an extra systematic
(H$\alpha$ and H$\beta$ fall on different humps); the qualitative
inversion, however, is robust and shared by all four targets.

\paragraph{Line widths}
The H$\alpha$ and H$\beta$ profiles are broad and single-peaked
(instrument-corrected full width at half maximum $\approx 350$--$1050$\,km\,s$^{-1}$, broadest
in LAMOST~J0035$+$4333), as expected for a magnetically channeled
accretion flow seen over a range of projected velocities; the narrow
(irradiated-donor) and broad (stream) components that phase-resolved
spectra of brighter polars separate \citep[e.g.,][]{2022MNRAS.511...20K}
are blended in these single-epoch co-adds.

Together with the cyclotron humps and the high and low light-curve states,
the emission-line diagnostics support magnetic accretion in all four
systems. The \citet{Silber1992} He\,II criterion independently confirms
this classification only for DESI~J0022$+$1340 and LAMOST~J0035$+$4333.

\subsection{Spectral energy distributions and the donors}
\label{sec:sed}

\begin{figure*}[ht]
\centering
\includegraphics[width=0.98\textwidth]{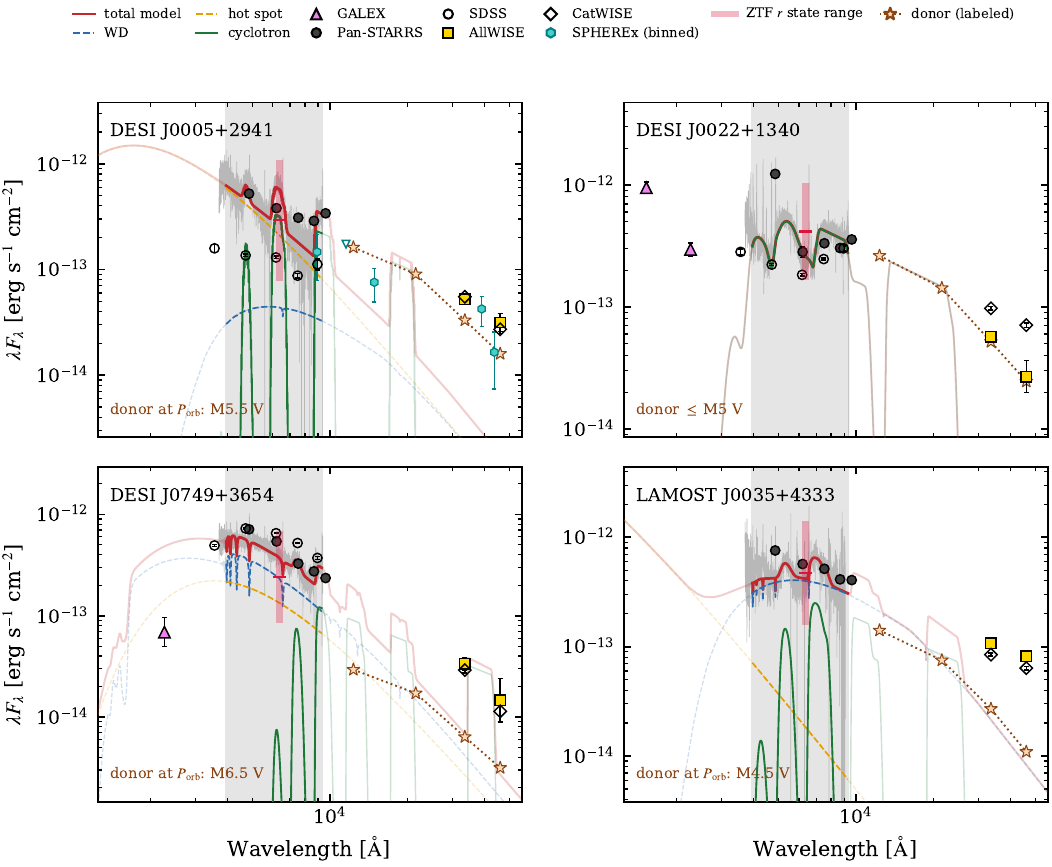}
\caption{Archival spectral energy distributions for DESI~J0005$+$2941
(\textit{top left}), DESI~J0022$+$1340 (\textit{top right}), DESI~J0749$+$3654 (\textit{bottom
left}), and LAMOST~J0035$+$4333 (\textit{bottom right}).  The
de-reddened survey spectrum is in gray.  Colored curves mark the joint fit of
Fig.~\ref{fig:joint_fits} and its components, solid over the fitted
3950--9300\,\AA{} window (gray band) and faint where extrapolated, the
low harmonics longward of 1\,$\mu$m being unconstrained by these spectra.
Symbols indicate the de-reddened archival photometry, with a separate symbol for each catalog because
the epochs span two decades: Pan-STARRS
(filled dark gray circles) and SDSS (open circles). The teal hexagons and downward triangle in
the DESI~J0005$+$2941 panel are the binned SPHEREx forced photometry and
a $2\sigma$ upper limit.  Red bars denote the 2nd--98th percentile range of the
ZTF $r$ light curve, i.e.,\ the accretion-state amplitude.  Brown stars denote
the donor at $J$, $K_{\rm s}$, $W1,$ and $W2$ -- the type expected from
the period--density relation except in DESI~J0022$+$1340, which has no
period; there the stars mark the most luminous donor the mid-infrared allows
(Sect.~\ref{sec:sed}).  The LAMOST spectrum has only a relative flux
scale, normalized to the mean ZTF $r$ magnitude, so its placement and the
donor limit drawn from it depend on the parallax distance of
Table~\ref{tab:sample}.
\label{fig:sed}}
\end{figure*}

Figure~\ref{fig:sed} places each fit in the context of the archival
photometry of Table~\ref{tab:sed}, from the far-UV (FUV) to
5\,$\mu$m. None of the four is detected by 2MASS, by the United Kingdom Infrared Telescope (UKIRT)
Hemisphere Survey, or by \textit{Spitzer}. All four fall inside the survey area of
SPHEREx (the Spectro-Photometer for the History of the Universe, Epoch of
Reionization and Ices Explorer), and we retrieved its forced photometry
for DESI~J0005$+$2941, but at $19.5$\,AB the source sits at the $5\sigma$
depth of the completed survey \citep{Bock2026}: the binned
spectrophotometry only reproduces the WISE flux ($55\pm18$ against
58\,$\mu$Jy) and adds nothing to it. The WISE $W1$ (3.35\,$\mu$m) and $W2$
(4.60\,$\mu$m) magnitudes therefore set the donor limits. None of this photometry is simultaneous with the spectra: the
ZTF light curves span 1.7--2.9\,mag (Sect.~\ref{sec:timing}), and the
catalogs record the same variability, with the Pan-STARRS and SDSS
magnitudes differing by up to 1.90\,mag and the AllWISE and CatWISE2020
magnitudes by up to 1.06\,mag, three to forty times the photometric
errors. The UV therefore does not constrain the WD,
which in a polar is outshone by the accretion-heated pole cap; in the
infrared we adopted the fainter of the two WISE epochs, which bounds the
donor whatever the accretion state.

Two independent constraints then bound the donors. A Roche-lobe-filling
donor has a mean density fixed by the orbital period, so
$R_2 \simeq 0.234\,(M_2/M_\odot)^{1/3}\,(P_{\rm orb}/{\rm hr})^{2/3}\,
R_\odot$ and stars earlier than some type are too large to fit;
independently, a donor of a given type expanded to that radius must not
outshine the observed $W1$. With the dwarf sequence of
\citet{PecautMamajek2013} the two together exclude donors earlier than about M4.5--M5\,V in
J0005$+$2941, J0022$+$1340 and J0749$+$3654, and about M3\,V in
J0035$+$4333; for J0022$+$1340, which has no measured period, only the
photometric constraint applies. The limits are approximate: the distance
uncertainty alone (7--16\%, and $\sim$40\% for J0035$+$4333) is worth
0.3--0.7\,mag in $M_{W1}$, about half a subtype on that sequence.

The same relation predicts the donors themselves -- M5.5\,V, M6.5\,V and
M4.5\,V at 105.3, 80.4 and 143.8\,min -- contributing 64\%, 22\% and
32\% of the observed $W1$. Only in J0005$+$2941 is the mid-infrared
therefore mostly photospheric, consistent with its $W1-W2 =
0.22\pm0.11$; the other three require an additional component. M-dwarf donors as a class are excluded nowhere in
the sample: the limits rule out only the early-M subtypes, with thresholds
between about M3 and M5\,V depending on the target, and are consistent with
the late-type donors the measured periods imply \citep{knigge2011}. None
of this enters the spectral fit, which has no donor component
(Sect.~\ref{sec:joint}), and none of it affects the fields, since a donor
adds a smooth red continuum rather than structure on the scale of the
cyclotron humps.

\section{Discussion}
\label{sec:discussion}

\subsection{Continuum placement and branch choice}
\label{sec:compare}

We used the two-step fit as a sensitivity test. A
smooth continuum was fitted in wavelength windows assumed to be free of
cyclotron emission; it was then subtracted, and the residual was fitted as a pure
cyclotron spectrum.
For DESI~J0005$+$2941, whose spectrum contains several well-separated
humps, our two-step test and the joint fit agree ($58.1$ vs.
$\BJaaa$\,MG), confirming that a well-resolved harmonic pattern fixes $B$
largely independently of the continuum treatment; the $45$\,MG catalog
value \citep{2026hou} corresponds to a different harmonic assignment that
is not favored by the present profiled scan (Sect.~\ref{sec:res_J0005}).

For the branch-ambiguous sources the result changes substantially.
J0022$+$1340 gives $44.6$\,MG (our two-step test), $64.4/77.7$\,MG
(joint), and $73$\,MG (the catalog pipeline; \citealp{2026hou}).
This comparison shows that, for broad or blended humps, different
continuum treatments can leave different harmonic branches acceptable.
J0749$+$3654 spans several tens of MG.
In these spectra the humps are broad or blended, so different continuum
placements leave different harmonic assignments acceptable. The main
uncertainty is then the branch choice, not the photon-noise error around
one field value. We therefore list the competing branches explicitly.

\subsection{Population context}
\label{sec:population}

The measured periods place the sample in the short-period regime.
DESI~J0749$+$3654 (80.4\,min) lies near the CV period minimum
\citep{1999kolb,knigge2011}. DESI~J0005$+$2941 (105.3\,min) is below
the period gap. LAMOST~J0035$+$4333 (143.8\,min) lies inside the gap,
where polars, unlike disk CVs, are not underrepresented
\citep{2021schreiber,2024schreiber}.
The field constraints are not equally strong. J0005$+$2941 has a secure
56.8\,MG field. J0022$+$1340 has two acceptable branches at 64.4 and
77.7\,MG. J0035$+$4333 has an indicative 51.1\,MG value from a
relative-flux, low-S/N spectrum. J0749$+$3654 remains widely branch
ambiguous.
These fields are typical of polars, but the sample is not representative:
it was selected because optical cyclotron humps are visible.
A hump-selected sample of four cannot constrain the field distribution or
its origin.

The donors fit the same picture. For the three systems with a measured
period the mid-infrared limits of Sect.~\ref{sec:sed} are consistent with
the late-type donors those periods imply, and for DESI~J0022$+$1340 the
photometric limit alone excludes types earlier than about M5\,V. The one
system where a deeper limit would matter is DESI~J0749$+$3654: at
80.4\,min its donor may be very late or already substellar
\citep{knigge2011,Cunningham2025}, and its expected $3.35\,\mu$m
contribution lies well below the present limit. Testing that needs
near-infrared photometry in a low state or a radial-velocity mass ratio.

\subsection{Accretion-region parameters}
\label{sec:physics}

With an absolute flux scale (DESI) and \textit{Gaia} distances, the fitted cyclotron
amplitudes can be converted into approximate luminosities and emitting
areas. We computed these only where the field is constrained well enough to
set a single physical scale: for J0005$+$2941, which has a unique field, and
for J0022$+$1340 at its formal lowest-$\chi^2$ branch as an illustrative
value; we do not quote them for J0749$+$3654, whose two acceptable branches
would imply different physical scales.
The cyclotron model peaks redward of the optical, so a bolometric
cyclotron luminosity requires extrapolating it beyond the fitted window.
We integrated the best-fit model over 3000--25000\,\AA{}; this range reaches
into the near-infrared, where the strong low harmonics -- down to the
fundamental near 1.5--1.9\,$\mu$m -- lie but the spectra give no direct
constraint. The result is therefore a model-dependent extrapolation, tied
to the indicative $kT$ and $\Lambda$ and best read as an order-of-magnitude estimate.
On this basis $L_{\rm cyc} \simeq \LcycJaaa$ and
$\LcycJbbb\,\mathrm{erg\,s^{-1}}$ for J0005$+$2941 and J0022$+$1340; the
geometric distances (uncertain at the 10--15\% level) add a further
$\sim$30\% normalization uncertainty to these luminosities and the rates
below.
Throughout we adopted a canonical WD,
$M_{\rm wd} = 0.6\,M_\odot$ and $R_{\rm wd} = 7\times10^{8}$\,cm; we did
not use the fitted WD solid angle for this calculation because it is
degenerate with the hot-spot continuum.
We estimated the projected emitting area by dividing the cyclotron
amplitude by the Rayleigh--Jeans peak intensity of the fitted
temperature \citep[the standard optically thick brightness estimate;
e.g.,][]{Cropper1990}. Expressed as a fraction of the WD
surface, this gives $f \simeq \fareaJaaa$ and $\fareaJbbb$. These values
fall in the $10^{-4}$--$10^{-2}$ range established for polar accretion
spots \citep{Cropper1990,wickramasinghe2000}.

The implied accretion rates, from the standard accretion-luminosity
relation $L_{\rm acc} = GM_{\rm wd}\dot M/R_{\rm wd}$
\citep[e.g.,][]{1995warner},
$\dot M \gtrsim L_{\rm cyc} R_{\rm wd}/GM_{\rm wd} \simeq
\MdotJaaa$--$\MdotJbbb\,M_\odot\,\mathrm{yr^{-1}}$,
are lower limits. Cyclotron emission is only one cooling channel, part
of the accretion luminosity is emitted in the X-ray and UV, and the DESI
spectra may not have been taken in the highest accretion state. Even if
these corrections raise the rates by an order of magnitude, the systems
remain in the low-accretion regime for polars
\citep[comparable to the low-accretion-rate systems of][]{Schmidt2005LowAccretion}.
In this regime, and at fields of tens of MG, cyclotron cooling can keep
the post-shock electron temperature low \citep{Fischer2001}. The low
$kT$ values in the optical fits are therefore expected and should not be
treated as direct measurements of the full shock temperature.

The magnetic moments of the well-constrained systems,
$\mu = B R_{\rm wd}^{3} \simeq \muJaaa$--$\muJbbb\,\mathrm{G\,cm^{3}}$,
are high enough to synchronize a short-period, low-accretion-rate system
\citep{1995warner,hellierbook}. This supports a polar interpretation for
J0005$+$2941. For J0022$+$1340 it makes a polar classification plausible,
but the missing spin and orbital periods still leave an intermediate
polar interpretation open.

\subsection{Systematic limitations}
\label{sec:systematics}

Our error bars assume a single emission region with one temperature and
one optical depth. Real accretion columns are not so uniform: their
temperature, density, and field vary across the emitting area, and a
single spectrum averages over all of it. Phase-resolved studies of bright
polars show that $kT$ and $\Lambda$ change substantially with viewing
phase, while $B$ is usually more stable
\citep{2008campbellb,2020kolbina,2022MNRAS.511...20K}. We therefore
interpret $B$ as the field of the dominant emitting region.

The benchmark recoveries give a $2.6\%$ method-level systematic on $B$
(the RMS deviation of EQ~Cet, BS~Tri, MQ~Dra, and PZ~Vir from their
published fields; Sect.~\ref{sec:benchmarks}, Appendix~\ref{app:sdss}). We added
this in quadrature to the formal error of every single-branch field. This
systematic is a floor because the benchmark spectra are cleaner than the
survey co-adds. When the humps are weak or blended, more than one harmonic number can
reproduce them, and each choice implies a different field (the same humps
might be, e.g., the $n=2$--4 or the $n=3$--5 harmonics). The dominant
uncertainty is then this discrete choice of branch, not the scatter
within a branch, so the $\pm$ errors we list for such branches measure
only the scatter within the branch.

A second limitation occurs when high $B$ and high $kT$ smear the low
cyclotron harmonics into a broad pseudo-continuum. Such a component can
fit the overall shape without reproducing the humps, as in
DESI~J0749$+$3654. We reduced this risk by requiring a structured
cyclotron component (contrast $>0.15$) and by fitting the hump-dominated
wavelength region. This is not the same as fixing the geometry by hand:
for BS~Tri a fit with a free $\theta$ returned $\approx86^\circ$, close to
the $87^\circ$ from its eclipse, so the angle is constrained by the data
rather than imposed, subject to the geometric caveat of
Sect.~\ref{sec:benchmarks}.

Independent geometric information remains valuable. An eclipse,
phase-resolved spectroscopy, polarimetry, or Zeeman features can break
degeneracies that a single spectrum cannot. DESI~J0005$+$2941 is the
most promising target in this sample because its ZTF light curve suggests
an eclipse.

Finally, the survey co-adds average over the 80--145\,min orbits. The
fitted viewing angle is therefore an effective value, not the instantaneous
geometry. Phase smearing mainly affects $kT$ and $\Lambda$ by broadening
the humps; the field, set by the hump positions, is less sensitive.

\subsection{Prospects}
\label{sec:prospects}

The method needs three common data products: a survey-quality spectrum,
broad-band time-domain photometry, and, when available, a parallax to set
the continuum scale. These now exist for many CV candidates from DESI,
LAMOST, SDSS-V, ZTF, and soon the Legacy Survey of Space and Time (LSST) of the Vera C. Rubin Observatory \citep{2023inighta,2025desidr1}.
The same workflow can therefore be applied to larger samples while
recording the harmonic ambiguity for each object.

Multi-epoch survey spectra would improve the constraints. Different
accretion states and viewing phases can separate true cyclotron humps
from continuum curvature and can reduce the $kT$--$\Lambda$ degeneracy.
At fields above $\sim$100\,MG, the low harmonics move into the
NUV and optical surveys miss part of the high-field tail.
NUV spectroscopy, for example with the Chinese Space Station
Telescope, would extend the same approach to that regime.

\section{Conclusions}
\label{sec:conclusions}

We have characterized four magnetic CV candidates from DESI DR1 and
LAMOST using a joint continuum--cyclotron forward model and 7\,yr of
ZTF photometry.
Our main results are:

\begin{enumerate}
\item DESI J0005$+$2941: a polar with
$B = \BJaaa\pm\BJaaaErr$\,MG (the error dominated by the $2.6\%$
benchmark systematic floor) and a secure $105.349\pm0.003$\,min
period whose folded light curve shows a $\geq$1.4\,mag, 2--4\,min
eclipse-like feature ($i \approx 77$--$80^\circ$): a strong
eclipsing-polar candidate below the period gap.
\item DESI J0022$+$1340: an almost purely cyclotron spectrum gives
a formal global minimum at $B = \BJbbb$\,MG with a nearly degenerate
alternative branch at 77.7\,MG, while its blue GALEX counterpart indicates
a hot UV component.
The subtype (polar vs.\ intermediate polar) is open, and the sparse ZTF
photometry yields no secure period.
\item DESI J0749$+$3654 (= SDSS J074917.11$+$365428.0): a known
polar near the period minimum whose single-epoch DESI spectrum is
branch-ambiguous (acceptable full-refit solutions near 29.3 and
96.4\,MG), so we do not quote a unique optical cyclotron field; we do,
however, refine its orbital period to $80.4062\pm0.0008$\,min from the
double-humped ZTF cyclotron-beaming light curve.
\item LAMOST J0035$+$4333: varies on a 143.8\,min period, or its
1-day alias at 159.7\,min, with polar-like state changes and a
coincident slew-survey X-ray source; its relative-flux, low-S/N spectra
give an indicative $B = \BJddd\pm\BJdddErr$\,MG that a flux-calibrated
spectrum could confirm.
\item On method: simultaneous continuum--cyclotron fitting gives a
reliable field when the humps are well resolved. When the humps are weak
or blended, the same fit gives the acceptable harmonic branches rather
than one unique field value.
The benchmark tests on EQ~Cet and BS~Tri, and the blind SDSS recoveries
of MQ~Dra and PZ~Vir (Appendix~\ref{app:sdss}), reproduce published
fields to within a few percent for well-constrained cases. The offset
branch recovered for J1344 illustrates the same harmonic-assignment
systematic seen in the branch-ambiguous targets.
Within a fixed branch, the field is the most robust output. The
WD and hot-spot continua are degenerate, so their individual
temperatures and flux ratios should not be over-interpreted; a free fit
can even place them at a combination no real WD occupies.
Tying the photosphere to a mass--radius relation and to the CV mass and
temperature ranges removes that and leaves the field unchanged, to
within 0.01\,MG at the adopted branch of each of the three DESI targets
(Sect.~\ref{sec:wdprior}).
\item Donors: the archival mid-infrared photometry excludes
donors earlier than about M4.5--M5\,V in J0005$+$2941, J0022$+$1340 and
J0749$+$3654 and about M3\,V in J0035$+$4333. For the three systems with
a measured period this is consistent with the late-type donors those
periods imply; J0022$+$1340 has no period and its limit is purely
photometric. Only in J0005$+$2941 is the 3.35\,$\mu$m flux mostly
the donor; the $W1-W2 = 0.68\pm0.05$ of J0035$+$4333 cannot be explained
by the donor alone and is likely accretion related
(Sect.~\ref{sec:sed}).
\end{enumerate}

Eclipsing polars below the period gap are rare; if its eclipse is
confirmed by high-speed photometry, J0005$+$2941 would yield the precise
component masses and inclination that make such systems valuable tests of
magnetic-CV evolution.

\section*{Data availability}
\label{sec:data_avail}

The DESI, LAMOST, ZTF, GALEX, 2MASS, AllWISE, and \textit{Gaia} data underlying
this work are available from the respective public archives.
The intermediate and final products of our analysis -- the de-reddened
binned spectra, the fitted model components for every panel of
Fig.~\ref{fig:joint_fits}, the profiled $\chi^2$ grids of
Figs.~\ref{fig:bench} and \ref{fig:B_profiles}, the benchmark residual
spectra and models, the MCMC chains, the light-curve analysis results,
the compiled archival photometry and SPHEREx spectrophotometry, and the
donor-limit calculation -- are distributed, together with the analysis
code,
in the project repository\footnote{\url{https://github.com/wangnengdejiamao/cyclotron-spectrum-analyzer}}.

\begin{acknowledgements}
J.L. and Chengyuan Li are supported by the National Natural Science
Foundation of China (NSFC Grant No. 12233013).
This work was also supported by the project ``Large-sample search for and study
of objects in the late stages of stellar evolution''
(Project No.~CMS-CSST-2025-A13).
We thank Alexander I.~Kolbin for kindly providing the phase-resolved
spectra of BS~Tri.
The EQ~Cet benchmark uses the phase-differenced spectrum of
\citet{2008campbellb}.

This work used public data from the Dark Energy Spectroscopic
Instrument (DESI), the Large Sky Area Multi-Object Fiber Spectroscopic
Telescope (LAMOST), the \textit{Zwicky} Transient Facility (ZTF), the ESA \textit{Gaia}
mission, and the GALEX, SDSS, Pan-STARRS, 2MASS, WISE, and SPHEREx
archives. This publication makes use of data products from the
Wide-field Infrared Survey Explorer and of the NASA/IPAC Infrared
Science Archive, which is funded by NASA and operated by the California
Institute of Technology. DESI is managed by
Lawrence Berkeley National Laboratory and supported by the U.S.
Department of Energy (Contract No. DE-AC02-05CH11231) and the U.S.
National Science Foundation; the DESI collaboration is honored to
conduct research on I'oligam Du'ag (Kitt Peak), a mountain with
particular significance to the Tohono O'odham Nation. LAMOST is funded
by the National Development and Reform Commission and operated by the
National Astronomical Observatories, Chinese Academy of Sciences. ZTF
is supported by the U.S. National Science Foundation (Grants
No. AST-1440341 and AST-2034437) and its partner institutions. \textit{Gaia}
data were processed by the \textit{Gaia} Data Processing and Analysis Consortium
(DPAC).

This research made use of numpy, scipy \citep{Virtanen2020}, astropy,
matplotlib, numba, emcee \citep{ForemanMackey2013}, and corner.
\end{acknowledgements}

\begin{appendix}
\onecolumn
\nolinenumbers
\section{MCMC sampling and posterior distributions}
\label{app:mcmc}

\begin{figure*}[ht]
\centering
\includegraphics[width=0.88\textwidth]{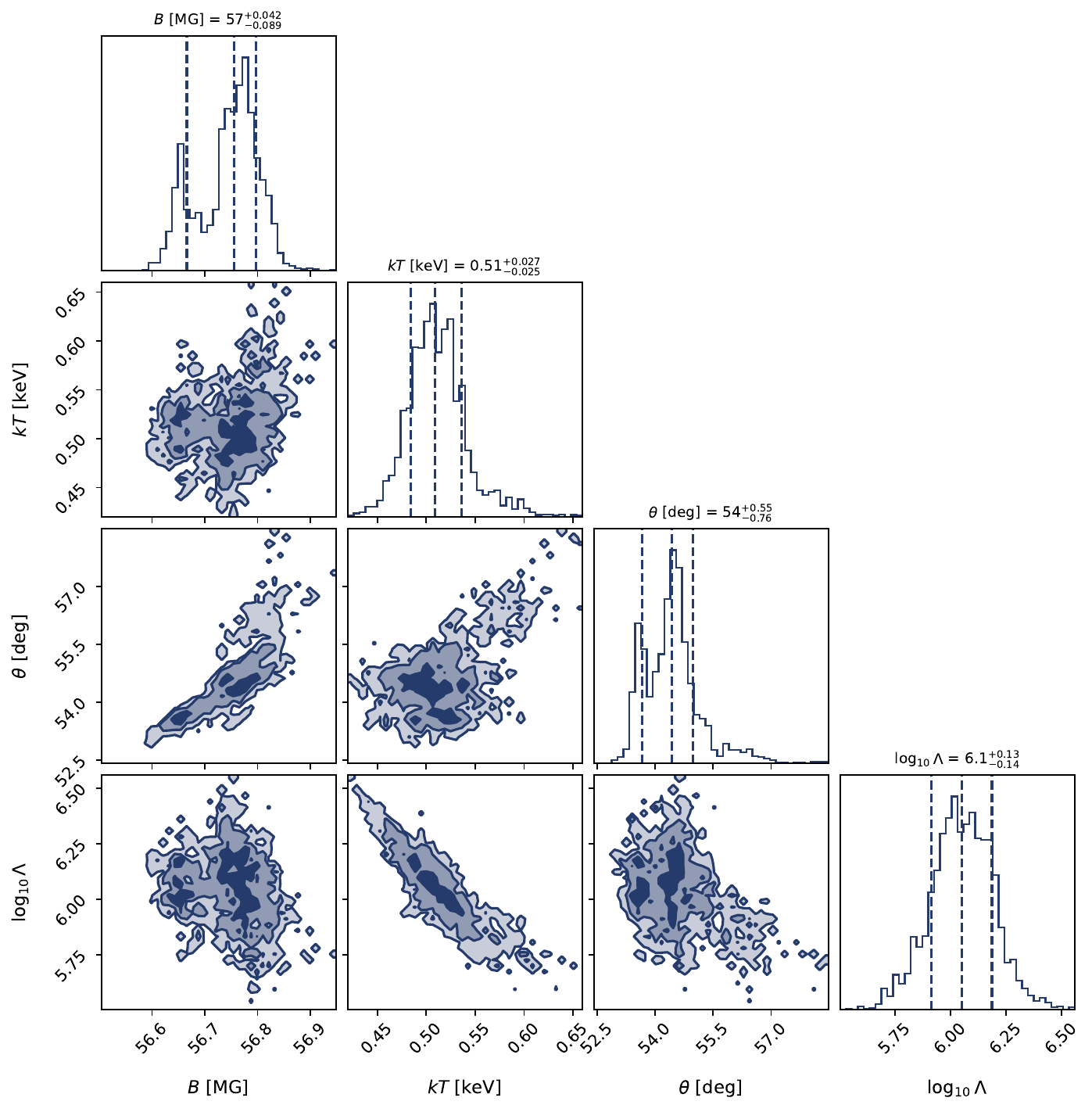}
\caption{Marginal and joint posterior distributions of the cyclotron
parameters $(B, kT, \theta, \log_{10}\Lambda)$ for DESI~J0005$+$2941
from the ten-parameter joint MCMC. Dashed lines mark the 16th, 50th,
and 84th percentiles.
\label{fig:corner}}
\end{figure*}

\begin{figure*}[ht]
\centering
\includegraphics[width=0.98\textwidth]{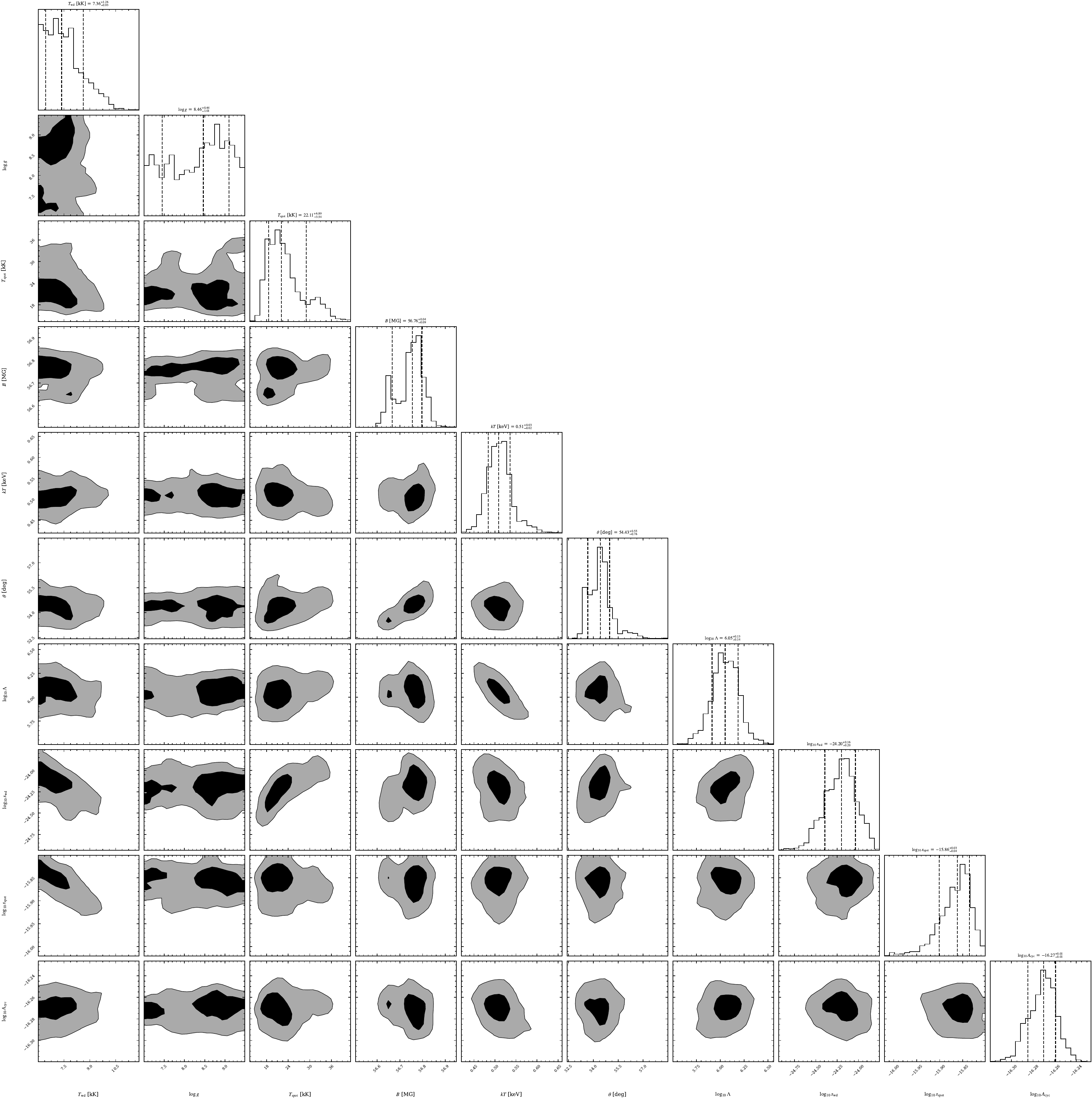}
\caption{All ten posterior distributions of the joint fit to
DESI~J0005$+$2941: the seven nonlinear parameters of
Eq.~(\ref{eq:model}) and the three component amplitudes.  Dashed lines
mark the 16th, 50th, and 84th percentiles.  The field is the only
tightly constrained parameter of the continuum-plus-cyclotron model;
$\log g$ is flat across its prior, and $T_{\rm spot}$, $s_{\rm wd}$ and
$s_{\rm spot}$ are mutually degenerate (Appendix~\ref{app:mcmc}).
\label{fig:corner_full}}
\end{figure*}

The ten parameters of Eq.~(\ref{eq:model}) are sampled with the
affine-invariant ensemble sampler \texttt{emcee}
\citep{ForemanMackey2013}, using 32 walkers run for 700 burn-in and
1800 production steps, with the log-likelihood tempered by
$\chi^2_{\rm min}/{\rm dof}$ (Sect.~\ref{sec:joint}).
The MCMC samples are local to the branch being quoted and therefore
measure branch-conditional uncertainties, not the relative probabilities
of separate harmonic assignments.
As a convergence check we split each production chain in half: for every
branch-specific run, the median cyclotron parameters of the two halves agree to within
0.5 of their $1\sigma$ widths (typically within $\sim0.3\sigma$, and
within $0.01$\,MG in $B$ for DESI~J0005$+$2941), and all marginal
posteriors are unimodal.
Figure~\ref{fig:corner} shows the marginal and joint posteriors of the
four cyclotron parameters for DESI~J0005$+$2941.
The field strength is tightly and unimodally constrained, whereas the
$kT$--$\Lambda$ degeneracy discussed in Sect.~\ref{sec:systematics}
appears as the correlated ridge; the nuisance continuum parameters
($T_{\rm wd}$, $\log g$, $T_{\rm hot}$, and the component
normalizations) are only weakly constrained. They are shown in
Fig.~\ref{fig:corner_full}, which displays the same chain over all ten
parameters. $B$ is constrained to $\pm0.07$\,MG and $kT$, $\theta$ and
$\log\Lambda$ form the usual correlated ridge, while the continuum
parameters are close to unconstrained: the 16th--84th percentile range of
$\log g$ spans 1.65 of the 2.49\,dex allowed by its prior, and
$T_{\rm spot}$ and $s_{\rm wd}$ are strongly anticorrelated, which is
the degeneracy of Sect.~\ref{sec:joint} seen directly. The posterior
medians ($T_{\rm wd} = 7363$\,K, $\log g = 8.46$,
$s_{\rm wd} = 10^{-24.20}$) imply $R_{\rm wd} = 0.021\,R_\odot$ and
$4.8\,M_\odot$, so the posterior of the free model is no more confined to
physical WDs than its $\chi^2$ minimum
(Sect.~\ref{sec:wdprior}).
DESI~J0005$+$2941 is shown as the well-constrained example; the full
ten-parameter chains for all targets are released with the data
products (Sect.~\ref{sec:data_avail}).

\FloatBarrier

\section{Blind validation on archival SDSS polars}
\label{app:sdss}

\begin{figure*}[ht]
\centering
\includegraphics[width=0.85\textwidth]{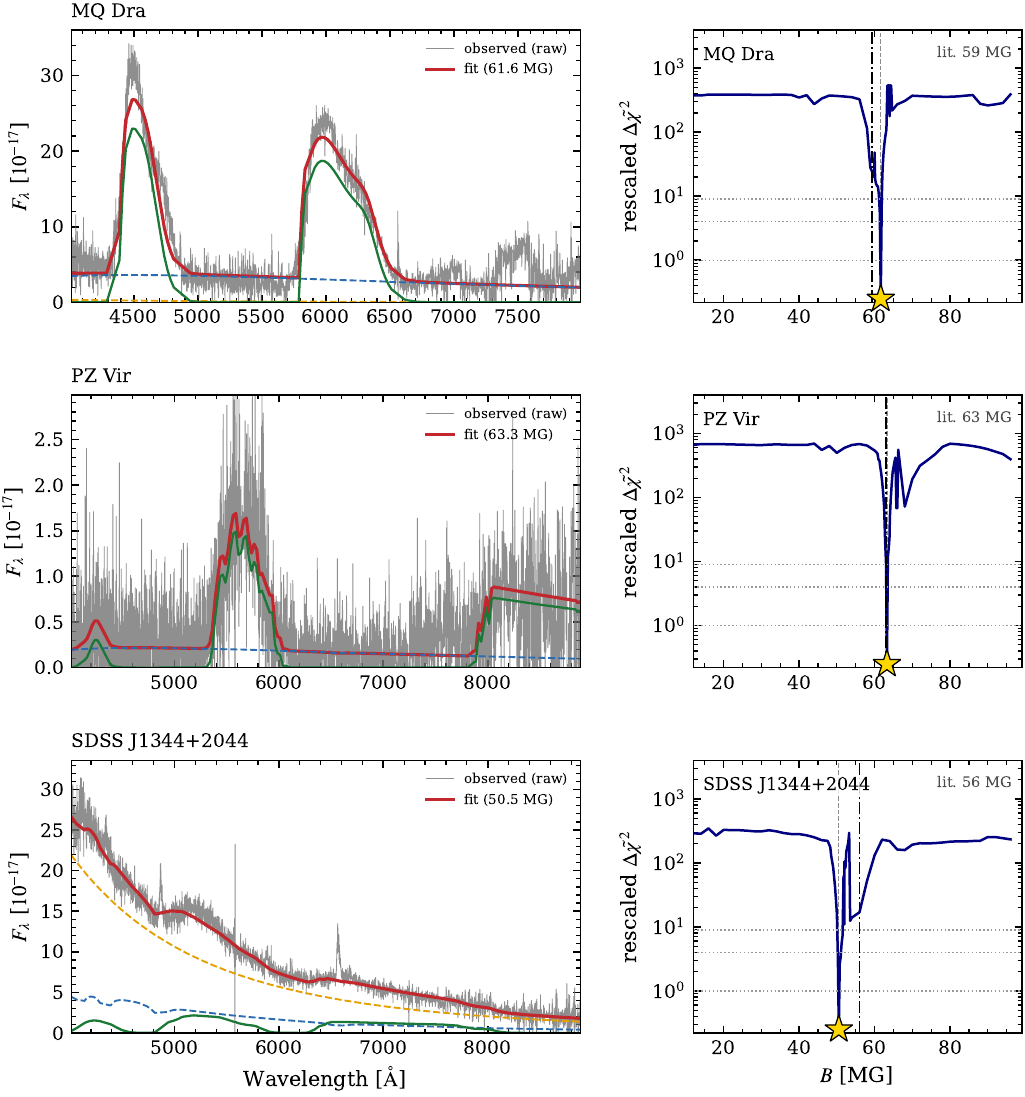}
\caption{Full-refit validation on archival SDSS spectra of three polars
with published fields.  \textit{Left}: Observed SDSS spectrum with the
total model and its WD, hot-spot blackbody, and cyclotron components.
\textit{Right}: Full-refit profiled $\chi^2(B)$ curves. The published
field is marked by a vertical dashed line and the recovered branch by a
star.  MQ~Dra and PZ~Vir are recovered close to their published fields;
J1344$+$2044 is recovered at a lower branch, illustrating the
harmonic-assignment systematic that also affects the science targets.
\label{fig:sdss_val}}
\end{figure*}

As a further check on real, single-epoch survey data, we applied the same
full-refit procedure to archival SDSS spectra of three polars with
independently published cyclotron fields: MQ~Dra
\citep{Szkody2003RareMagneticCVs}, PZ~Vir
\citep{Schmidt2005LowAccretion}, and
SDSS~J134441.83$+$204408.3 \citep{Littlefield2023J1344}.  For MQ~Dra a
red M-dwarf contribution was removed with an SDSS M-dwarf template
\citep{2007AJ....133..531B} before fitting.
The test is intentionally reported as a consistency check, not as proof
of universal precision: as Table~\ref{tab:sdss_val} and
Fig.~\ref{fig:sdss_val} show, MQ~Dra and PZ~Vir are recovered near their
published fields, while J1344$+$2044 is recovered at a lower branch.
This behavior is exactly the systematic relevant to the science
targets.

\begin{table}[ht]
\caption{Full-refit recovery of published fields from archival SDSS spectra.}
\label{tab:sdss_val}
\centering
\begin{tabular}{lcc}
\hline\hline
Polar & Published $B$ & Recovered $B$ \\
      & (MG) & (MG) \\
\hline
MQ~Dra & $\approx$59 & 61.6 \\
PZ~Vir & $\approx$63 & 63.3 \\
SDSS~J1344$+$2044 & $56\pm2$ & 50.5 \\
\hline
\end{tabular}
\tablefoot{Published fields are from
\citet{Szkody2003RareMagneticCVs} for MQ~Dra,
\citet{Schmidt2005LowAccretion} for PZ~Vir, and
\citet{Littlefield2023J1344} for SDSS~J1344$+$2044.
The J1344$+$2044 offset illustrates that sparse or low-contrast hump
patterns can select a neighboring harmonic branch.}
\end{table}

\FloatBarrier

\section{Alternative branch decompositions}
\label{app:branches}

\begin{figure*}[ht]
\centering
\includegraphics[width=0.98\textwidth]{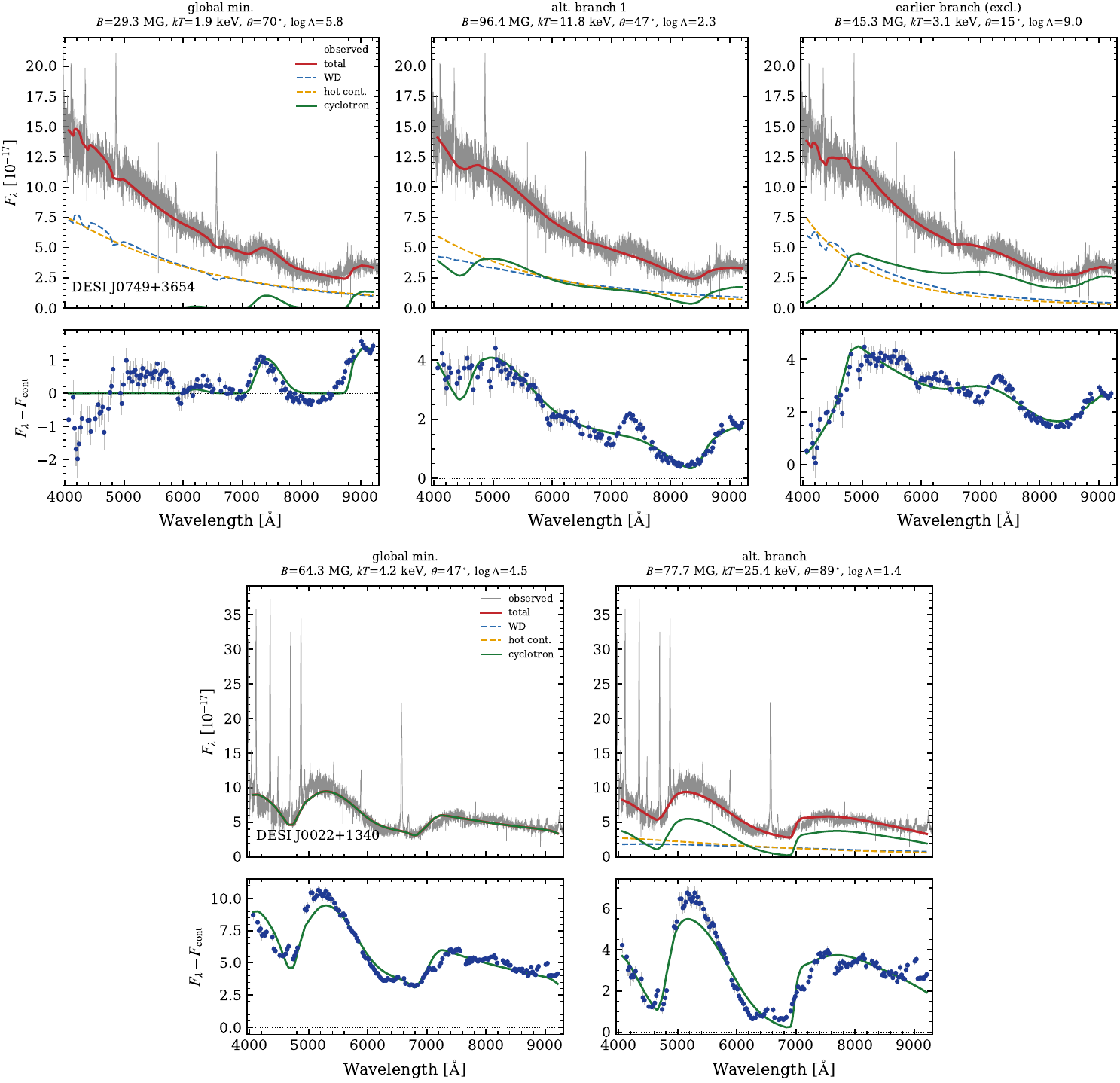}
\caption{Competing harmonic branches for the two branch-ambiguous targets,
in the same decomposition style as Fig.~\ref{fig:joint_fits}.
\textit{Top}: J0749$+$3654.  The 29.3 and 96.4\,MG branches have lower formal
$\chi^2$ than the 45.3\,MG comparison branch, but the decompositions show that the
continuum--cyclotron degeneracy remains severe, so the field is not quoted
as unique.
\textit{Bottom}: J0022$+$1340.  The 64.4\,MG branch is the formal global
minimum, but the 77.7\,MG branch fits the broad cyclotron humps nearly as
well, so the field is branch-ambiguous.
\label{fig:branches_ambig}}
\end{figure*}

For the two branch-ambiguous science targets (J0022$+$1340 and
J0749$+$3654) Fig.~\ref{fig:branches_ambig} plots the competing
decompositions using the same model components and likelihood as in
Fig.~\ref{fig:joint_fits}.
LAMOST~J0035$+$4333 is not shown here: its second sub-minimum
(54\,MG; Fig.~\ref{fig:B_profiles}) is the same harmonic at a different
$kT$, not a competing branch, and its adopted fit already appears in
Fig.~\ref{fig:joint_fits}.

\FloatBarrier

\section{Fits with a physically constrained white dwarf}
\label{app:wdprior}

\begin{figure*}[ht]
\centering
\includegraphics[width=0.98\textwidth]{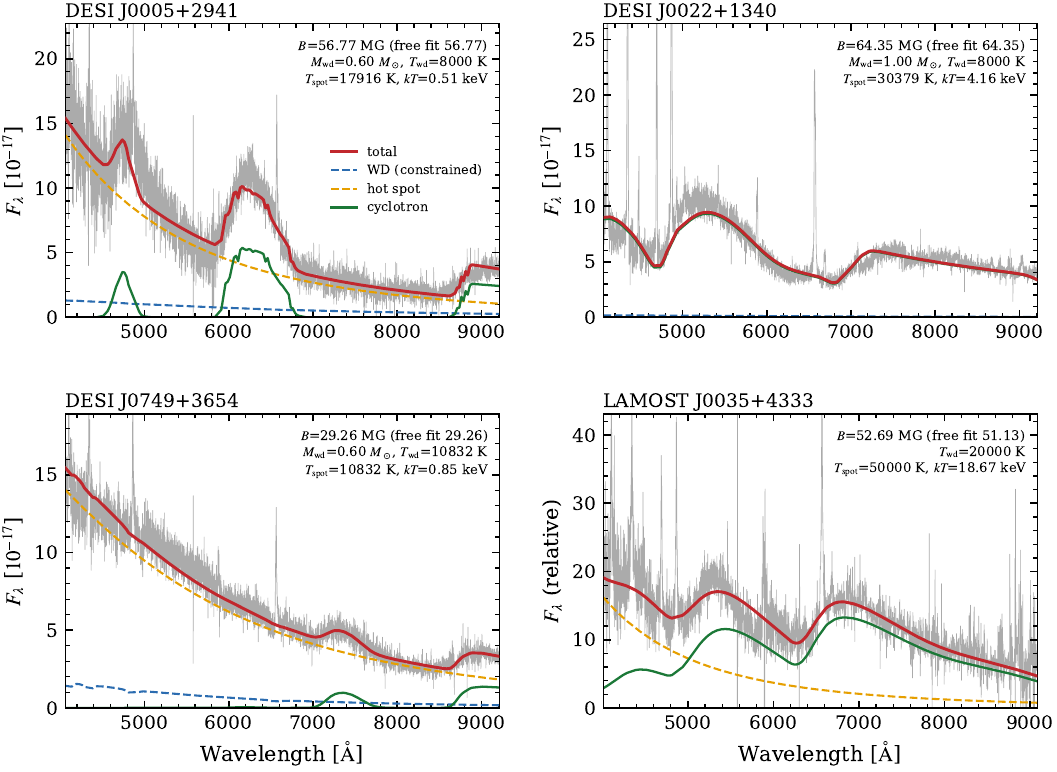}
\caption{Joint fits with the WD component tied to a
mass--radius relation for DESI~J0005$+$2941
(\textit{top left}), DESI~J0022$+$1340 (\textit{top right}), DESI~J0749$+$3654 (\textit{bottom
left}), and LAMOST~J0035$+$4333 (\textit{bottom right}). Shown are the total model (red), constrained WD
photosphere (dashed blue), hot spot (dashed orange), and cyclotron (green) components over the de-reddened spectrum (gray).  Each panel
gives the constrained field with the free-fit value in parentheses, the
WD mass and temperature, the hot-spot temperature, and the
electron temperature.
\label{fig:wdprior}}
\end{figure*}

Figure~\ref{fig:wdprior} shows the four spectra refitted with the
photosphere re-parametrized by its mass (Sect.~\ref{sec:wdprior};
Table~\ref{tab:continuum}). It now contributes 2--11\% of the optical
flux of the three DESI targets, as expected for a $\sim$10$^4$\,K WD at these distances, rather than the unphysical combinations listed
in Table~\ref{tab:continuum}, and the field is unchanged: flux moves
between the photosphere and the hot spot, while the harmonic spacing that
fixes $B$ is insensitive to how that flux is divided.
For DESI~J0022$+$1340 both continuum components remain negligible, the
spectrum being cyclotron dominated, and for LAMOST~J0035$+$4333,
with only the temperature prior available, both temperatures run to
their bounds and the field shifts by 1.6\,MG, within the ridge its
profile already shows (Sect.~\ref{sec:res_J0035}).

\FloatBarrier

\section{Emission-line spectra}
\label{app:emission}

\begin{figure*}[ht]
\centering
\includegraphics[width=0.90\textwidth]{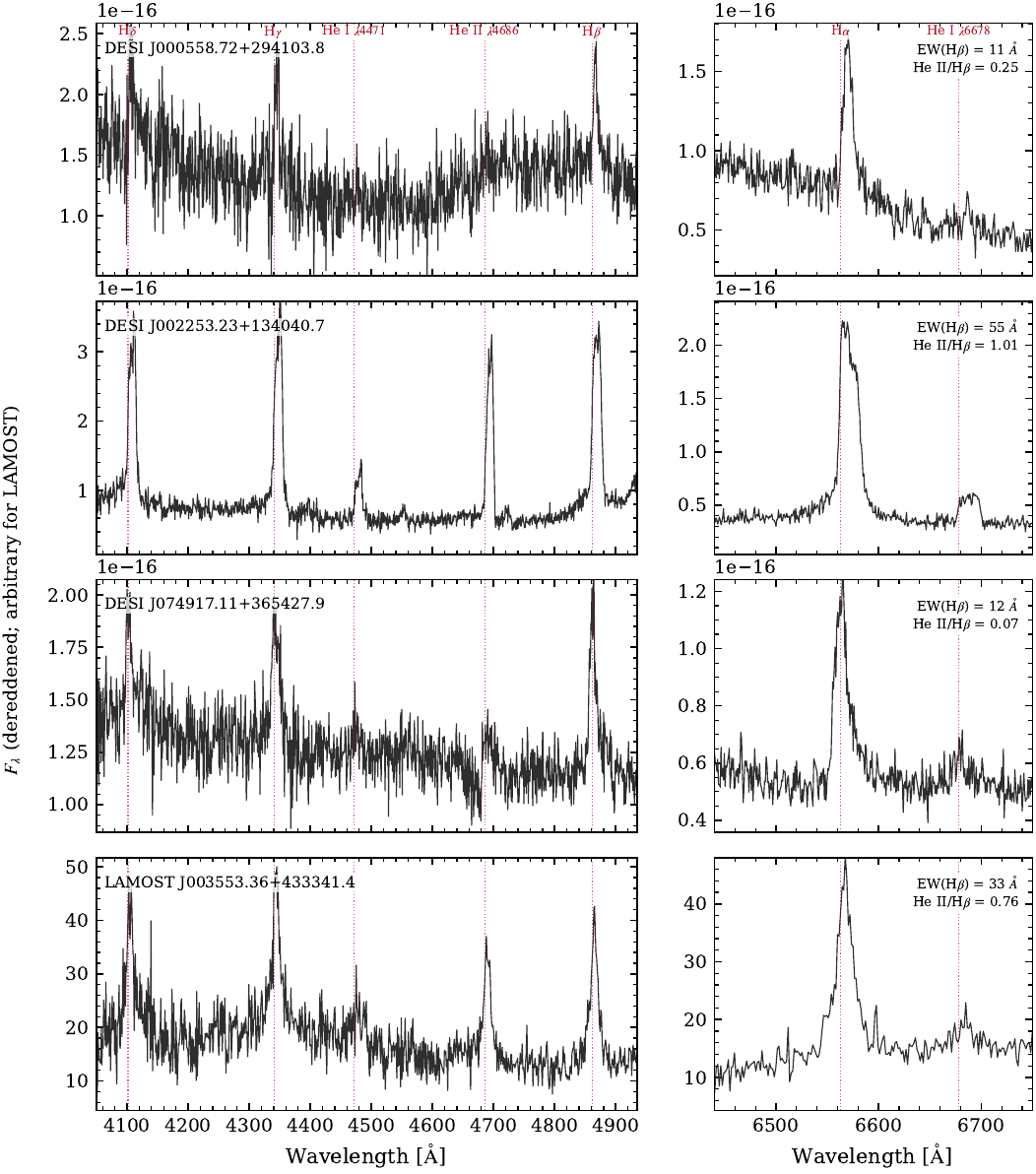}
\caption{Emission-line regions of the four targets (de-reddened spectra).
\textit{Left}: H$\delta$--H$\beta$ region, which contains the
high-excitation He\,II~$\lambda$4686 line. \textit{Right}: H$\alpha$ region.
Measured lines are marked (dotted), with line identifications labeled
in the upper panels, and the He\,II/H$\beta$ EW ratio and EW(H$\beta$)
are annotated.
The strong He\,II of DESI~J0022$+$1340 and LAMOST~J0035$+$4333 and
the flat Balmer decrements of all four are consistent with an accretion-flow origin
for the lines (Sect.~\ref{sec:emission}).
\label{fig:emission}}
\end{figure*}

Figure~\ref{fig:emission} shows the spectral regions from which the
EWs and fluxes of Table~\ref{tab:emission} were measured
(Sect.~\ref{sec:emission}).

\begin{table*}[ht]
\caption{Optical emission-line measurements.\label{tab:emission}}
\centering
\begin{tabular}{lcccccc}
\hline\hline
Source & EW(H$\alpha$) & EW(H$\beta$) & EW(He\,II) &
He\,II/H$\beta$ & H$\alpha$/H$\beta$ & H$\gamma$/H$\beta$ \\
 & (\AA) & (\AA) & (\AA) & & & \\
\hline
DESI~J0005$+$2941 & $18\pm1$ & $11\pm1$ & $3\pm1$ & $0.25\pm0.11$ & $0.94\pm0.13$ & $0.91\pm0.19$ \\
DESI~J0022$+$1340 & $100\pm2$ & $55\pm1$ & $55\pm2$ & $1.01\pm0.04$ & $0.92\pm0.02$ & $1.00\pm0.03$ \\
DESI~J0749$+$3654 & $16\pm1$ & $12\pm1$ & $1\pm1$ & $0.07\pm0.05$ & $0.67\pm0.06$ & $0.75\pm0.09$ \\
LAMOST~J0035$+$4333 & $46\pm3$ & $33\pm3$ & $25\pm3$ & $0.76\pm0.11$ & $1.42\pm0.16$ & $0.87\pm0.12$ \\
\hline
\end{tabular}
\tablefoot{Equivalent widths are emission-positive, measured by
local-continuum integration; uncertainties are Monte-Carlo estimates
from the pixel errors. He\,II denotes the $\lambda$4686 line.
The Balmer decrements use line fluxes for the flux-calibrated DESI
spectra and EWs for the relative-flux LAMOST spectrum; line widths are
discussed in Sect.~\ref{sec:emission}.
DESI~J0022$+$1340 and LAMOST~J0035$+$4333 satisfy the \citet{Silber1992}
magnetic criterion.}
\end{table*}

\clearpage

\section{Archival photometry}
\label{app:photometry}

Table~\ref{tab:sed} lists the archival photometry plotted in
Fig.~\ref{fig:sed} and used for the donor limits of
Sect.~\ref{sec:sed}.

\begin{table*}[ht]
\caption{Archival photometry of the four targets.\label{tab:sed}}
\centering
\begin{tabular}{llcccc}
\hline\hline
Band & $\lambda_{\rm eff}$ & DESI\,J0005$+$2941 & DESI\,J0022$+$1340 & DESI\,J0749$+$3654 & LAMOST\,J0035$+$4333 \\
 & (\AA) & (mag) & (mag) & (mag) & (mag) \\
\hline
GALEX FUV & 1528 & \ldots & $19.69\pm0.12$ & \ldots & \ldots \\
GALEX NUV & 2271 & \ldots & $20.52\pm0.13$ & $22.11\pm0.36$ & \ldots \\
SDSS $u$ & 3557 & $20.92\pm0.09$ & $20.42\pm0.06$ & $19.75\pm0.03$ & \ldots \\
SDSS $g$ & 4702 & $20.73\pm0.03$ & $20.30\pm0.02$ & $18.96\pm0.01$ & \ldots \\
PS1 $g$ & 4849 & $19.23\pm0.06$ & $18.40\pm0.04$ & $18.95\pm0.04$ & $18.89\pm0.08$ \\
SDSS $r$ & 6175 & $20.44\pm0.03$ & $20.14\pm0.03$ & $18.73\pm0.01$ & \ldots \\
PS1 $r$ & 6201 & $19.27\pm0.04$ & $19.66\pm0.11$ & $18.93\pm0.10$ & $18.88\pm0.05$ \\
SDSS $i$ & 7490 & $20.64\pm0.05$ & $19.57\pm0.02$ & $18.73\pm0.01$ & \ldots \\
PS1 $i$ & 7535 & $19.26\pm0.03$ & $19.23\pm0.04$ & $19.23\pm0.09$ & $18.74\pm0.01$ \\
PS1 $z$ & 8674 & $19.16\pm0.05$ & $19.15\pm0.07$ & $19.24\pm0.06$ & $18.80\pm0.05$ \\
SDSS $z$ & 8946 & $20.15\pm0.13$ & $19.11\pm0.05$ & $18.87\pm0.04$ & \ldots \\
PS1 $y$ & 9628 & $18.86\pm0.08$ & $18.84\pm0.07$ & $19.28\pm0.08$ & $18.69\pm0.03$ \\
AllWISE $W1$ & 33526 & $16.81\pm0.09$ & $16.72\pm0.10$ & $17.30\pm0.15$ & $16.04\pm0.05$ \\
CatWISE $W1$ & 33526 & $16.76\pm0.04$ & $16.14\pm0.03$ & $17.45\pm0.08$ & $16.30\pm0.03$ \\
AllWISE $W2$ & 46028 & $16.38\pm0.21$ & $16.56\pm0.33$ & $17.21\pm0.53$ & $15.35\pm0.08$ \\
CatWISE $W2$ & 46028 & $16.54\pm0.10$ & $15.50\pm0.05$ & $17.49\pm0.26$ & $15.62\pm0.04$ \\
\hline
\end{tabular}
\tablefoot{Sources: GALEX \citep{Bianchi2017}, SDSS~DR16 \citep{Ahumada2020}, Pan-STARRS\,1 \citep[PS1;][]{Chambers2016}, 2MASS \citep{Skrutskie2006}, the United Kingdom Infrared Telescope (UKIRT) Hemisphere Survey, the AllWISE and CatWISE2020 catalogs of the Wide-field Infrared Survey Explorer \citep[WISE;][]{Wright2010,Marocco2021}, and the \textit{Spitzer} Enhanced Imaging Products. Magnitudes are as cataloged (AB for GALEX, SDSS and Pan-STARRS, Vega for WISE) and are not de-reddened; the SED of Fig.~\ref{fig:sed} applies the extinction of Table~\ref{tab:sample}. None of the four targets is detected by 2MASS, by the UKIRT Hemisphere Survey or by \textit{Spitzer}. The AllWISE $W3$ and $W4$ entries are upper limits in all four cases and are omitted. The two independent WISE reductions (AllWISE, epochs 2010--2011; CatWISE2020, epochs 2010--2018) and the two independent optical surveys are listed separately because their differences measure the accretion-state variability rather than the photometric error (Sect.~\ref{sec:sed}).}
\end{table*}

\end{appendix}


\begin{thebibliography}{74}
\expandafter\ifx\csname natexlab\endcsname\relax\def\natexlab#1{#1}\fi

\bibitem[{{Ahumada} {et~al.}(2020){Ahumada}, {Allende Prieto}, {Almeida},
  {et~al.}}]{Ahumada2020}
{Ahumada}, R., {Allende Prieto}, C., {Almeida}, A., {et~al.} 2020, \apjs, 249,
  3

\bibitem[{{Bailer-Jones} {et~al.}(2021){Bailer-Jones}, {Rybizki}, {Fouesneau},
  {Demleitner}, \& {Andrae}}]{BailerJones2021}
{Bailer-Jones}, C.~A.~L., {Rybizki}, J., {Fouesneau}, M., {Demleitner}, M., \&
  {Andrae}, R. 2021, \aj, 161, 147

\bibitem[{{Beard} \& {Baker}(1962)}]{1962BB}
{Beard}, D.~B. \& {Baker}, J.~C. 1962, Physics of Fluids, 5, 1113

\bibitem[{{Bellm} {et~al.}(2019){Bellm}, {Kulkarni}, {Graham}, \&
  et~al.}]{Bellm2019}
{Bellm}, E.~C., {Kulkarni}, S.~R., {Graham}, M.~J., \& et~al. 2019, \pasp, 131,
  018002

\bibitem[{Belloni {et~al.}(2020)Belloni, Schreiber, Pala, G{\"a}nsicke,
  Zorotovic, \& Rodrigues}]{2020belloni}
Belloni, D., Schreiber, M.~R., Pala, A.~F., {et~al.} 2020, \mnras, 491, 5717

\bibitem[{{Bianchi} {et~al.}(2017){Bianchi}, {Shiao}, \&
  {Thilker}}]{Bianchi2017}
{Bianchi}, L., {Shiao}, B., \& {Thilker}, D. 2017, \apjs, 230, 24

\bibitem[{{Bochanski} {et~al.}(2007){Bochanski}, {West}, {Hawley}, \&
  {Covey}}]{2007AJ....133..531B}
{Bochanski}, J.~J., {West}, A.~A., {Hawley}, S.~L., \& {Covey}, K.~R. 2007,
  \aj, 133, 531

\bibitem[{{Bock} {et~al.}(2026){Bock}, {Aboobaker}, {Adamo},
  {et~al.}}]{Bock2026}
{Bock}, J.~J., {Aboobaker}, A.~M., {Adamo}, J., {et~al.} 2026, \apj, 999, 139

\bibitem[{Briggs {et~al.}(2018)Briggs, Ferrario, Tout, \&
  Wickramasinghe}]{2018briggs}
Briggs, G.~P., Ferrario, L., Tout, C.~A., \& Wickramasinghe, D.~T. 2018,
  \mnras, 481, 3604

\bibitem[{Campbell {et~al.}(2008)Campbell, Harrison, Mason, Howell, \&
  Schwope}]{2008campbellb}
Campbell, R.~K., Harrison, T.~E., Mason, E., Howell, S., \& Schwope, A.~D.
  2008, \apj, 678, 1304

\bibitem[{{Cardelli} {et~al.}(1989){Cardelli}, {Clayton}, \&
  {Mathis}}]{CCM1989}
{Cardelli}, J.~A., {Clayton}, G.~C., \& {Mathis}, J.~S. 1989, \apj, 345, 245

\bibitem[{{Chambers} {et~al.}(2016){Chambers}, {Magnier}, {Metcalfe},
  {et~al.}}]{Chambers2016}
{Chambers}, K.~C., {Magnier}, E.~A., {Metcalfe}, N., {et~al.} 2016, arXiv
  e-prints, arXiv:1612.05560

\bibitem[{{Chanmugam}(1980)}]{1980chan}
{Chanmugam}, G. 1980, \apj, 241, 1122

\bibitem[{{Chanmugam} \& {Dulk}(1981)}]{1981chanmugam}
{Chanmugam}, G. \& {Dulk}, G.~A. 1981, \apj, 244, 569

\bibitem[{Cropper(1990)}]{Cropper1990}
Cropper, M. 1990, \ssr, 54, 195

\bibitem[{{Cui} {et~al.}(2012){Cui}, {Zhao}, {Chu}, \& et~al.}]{Cui2012}
{Cui}, X.-Q., {Zhao}, Y.-H., {Chu}, Y.-Q., \& et~al. 2012, Research in
  Astronomy and Astrophysics, 12, 1197

\bibitem[{{Cunningham} {et~al.}(2025){Cunningham}, {Caiazzo}, {Sienkiewicz},
  {Wheatley}, {G{\"a}nsicke}, {El-Badry}, {Arcodia}, {Charbonneau}, {Connor},
  {De}, {Hakala}, {Kenyon}, {Maheshwari}, {Rodriguez}, {van Roestel}, \&
  {Tremblay}}]{Cunningham2025}
{Cunningham}, T., {Caiazzo}, I., {Sienkiewicz}, G., {et~al.} 2025, \mnras, 540,
  633

\bibitem[{{Denisenko} {et~al.}(2015){Denisenko}, {Lipunov}, {Gorbovskoy}, \&
  et~al.}]{ATel6913}
{Denisenko}, D., {Lipunov}, V., {Gorbovskoy}, E., \& et~al. 2015, {MASTER OT
  J002253.29+134041.0 discovery - new CV}, The Astronomer's Telegram, No. 6913

\bibitem[{{DESI Collaboration}(2025)}]{2025desidr1}
{DESI Collaboration}. 2025, arXiv e-prints [\eprint[arXiv]{2503.14745}]

\bibitem[{{Echevarr{\'\i}a}(1988)}]{Echevarria1988}
{Echevarr{\'\i}a}, J. 1988, \mnras, 233, 513

\bibitem[{Ferrario {et~al.}(2015)Ferrario, {de Martino}, \&
  G{\"a}nsicke}]{2015ferrario}
Ferrario, L., {de Martino}, D., \& G{\"a}nsicke, B.~T. 2015, \ssr, 191, 111

\bibitem[{{Fischer} \& {Beuermann}(2001)}]{Fischer2001}
{Fischer}, A. \& {Beuermann}, K. 2001, \aap, 373, 211

\bibitem[{Foreman-Mackey {et~al.}(2013)Foreman-Mackey, Hogg, Lang, \&
  Goodman}]{ForemanMackey2013}
Foreman-Mackey, D., Hogg, D.~W., Lang, D., \& Goodman, J. 2013, \pasp, 125, 306

\bibitem[{{Gaia Collaboration} {et~al.}(2023){Gaia Collaboration}, {Vallenari},
  {Brown}, \& et~al.}]{GaiaDR3}
{Gaia Collaboration}, {Vallenari}, A., {Brown}, A.~G.~A., \& et~al. 2023, \aap,
  674, A1

\bibitem[{{Gentile Fusillo} {et~al.}(2019){Gentile Fusillo}, {Tremblay},
  {G{\"a}nsicke}, \& et~al.}]{GentileFusillo2019}
{Gentile Fusillo}, N.~P., {Tremblay}, P.-E., {G{\"a}nsicke}, B.~T., \& et~al.
  2019, \mnras, 482, 4570

\bibitem[{{Ginzburg} \& {Gurevich}(1960)}]{1960SvPhU...3..115G}
{Ginzburg}, V.~L. \& {Gurevich}, A.~V. 1960, Soviet Physics Uspekhi, 3, 115

\bibitem[{{Hellier}(2001)}]{hellierbook}
{Hellier}, C. 2001, {Cataclysmic Variable Stars} (London: Springer-Praxis)

\bibitem[{{Hou} {et~al.}(2026){Hou}, {Lin}, {Luo}, {Dai}, {Dong}, \&
  {Ma}}]{2026hou}
{Hou}, W., {Lin}, J.-M., {Luo}, A.-L., {et~al.} 2026, \apjs, 282, 26

\bibitem[{Inight {et~al.}(2023)Inight, G{\"a}nsicke, Breedt, Israel,
  Littlefair, Manser, Marsh, Mulvany, Pala, \& Thorstensen}]{2023inighta}
Inight, K., G{\"a}nsicke, B.~T., Breedt, E., {et~al.} 2023, \mnras, 524, 4867

\bibitem[{Knigge {et~al.}(2011)Knigge, Baraffe, \& Patterson}]{knigge2011}
Knigge, C., Baraffe, I., \& Patterson, J. 2011, \apjs, 194, 28

\bibitem[{Koester(2010)}]{Koester2010}
Koester, D. 2010, Memorie della Societa Astronomica Italiana, 81, 921

\bibitem[{Kolb \& Baraffe(1999)}]{1999kolb}
Kolb, U. \& Baraffe, I. 1999, \mnras, 309, 1034

\bibitem[{Kolbin \& Borisov(2020)}]{2020kolbina}
Kolbin, A.~I. \& Borisov, N.~V. 2020, Astronomy Letters, 46, 812

\bibitem[{{Kolbin} {et~al.}(2022){Kolbin}, {Borisov}, {Serebriakova},
  {Shimansky}, {Katysheva}, {Gabdeev}, \& {Shugarov}}]{2022MNRAS.511...20K}
{Kolbin}, A.~I., {Borisov}, N.~V., {Serebriakova}, N.~A., {et~al.} 2022,
  \mnras, 511, 20

\bibitem[{Lin {et~al.}(2025)Lin, Ren, Li, Nancy, Cang, Ge, Tam, Huang, Li,
  Wang, Huang, \& Ma}]{2025lin}
Lin, J., Ren, L., Li, C., {et~al.} 2025, \aap, 694, A112

\bibitem[{{Littlefield} {et~al.}(2018){Littlefield}, {Garnavich}, {Hoyt}, \&
  {Kennedy}}]{littlefield2018}
{Littlefield}, C., {Garnavich}, P., {Hoyt}, T.~J., \& {Kennedy}, M. 2018, \aj,
  155, 18

\bibitem[{{Littlefield} {et~al.}(2023){Littlefield}, {Mason}, {Garnavich},
  {et~al.}}]{Littlefield2023J1344}
{Littlefield}, C., {Mason}, P.~A., {Garnavich}, P., {et~al.} 2023, \apjl, 943,
  L24

\bibitem[{{Lomb}(1976)}]{Lomb1976}
{Lomb}, N.~R. 1976, \apss, 39, 447

\bibitem[{{Ma{\'\i}z Apell{\'a}niz}(2022)}]{MaizApellaniz2022}
{Ma{\'\i}z Apell{\'a}niz}, J. 2022, \aap, 657, A130

\bibitem[{{Marocco} {et~al.}(2021){Marocco}, {Eisenhardt}, {Fowler},
  {et~al.}}]{Marocco2021}
{Marocco}, F., {Eisenhardt}, P.~R.~M., {Fowler}, J.~W., {et~al.} 2021, \apjs,
  253, 8

\bibitem[{{Masci} {et~al.}(2019){Masci}, {Laher}, {Rusholme}, \&
  et~al.}]{Masci2019}
{Masci}, F.~J., {Laher}, R.~R., {Rusholme}, B., \& et~al. 2019, \pasp, 131,
  018003

\bibitem[{{Meggitt} \& {Wickramasinghe}(1982)}]{1982MW}
{Meggitt}, S.~M.~A. \& {Wickramasinghe}, D.~T. 1982, \mnras, 198, 71

\bibitem[{{Nauenberg}(1972)}]{Nauenberg1972}
{Nauenberg}, M. 1972, \apj, 175, 417

\bibitem[{{Pala} {et~al.}(2022){Pala}, {G{\"a}nsicke}, {Belloni},
  {et~al.}}]{Pala2022}
{Pala}, A.~F., {G{\"a}nsicke}, B.~T., {Belloni}, D., {et~al.} 2022, \mnras,
  510, 6110

\bibitem[{Pala {et~al.}(2020)Pala, G{\"a}nsicke, Breedt, Knigge, Hermes,
  Gentile~Fusillo, Hollands, Naylor, Pelisoli, Schreiber, Toonen, Aungwerojwit,
  Cukanovaite, Dennihy, Manser, Pretorius, Scaringi, \& Toloza}]{2020pala}
Pala, A.~F., G{\"a}nsicke, B.~T., Breedt, E., {et~al.} 2020, \mnras, 494, 3799

\bibitem[{{Pecaut} \& {Mamajek}(2013)}]{PecautMamajek2013}
{Pecaut}, M.~J. \& {Mamajek}, E.~E. 2013, \apjs, 208, 9

\bibitem[{{Rodriguez} {et~al.}(2023){Rodriguez}, {Kulkarni}, {Prince},
  {Szkody}, {Burdge}, {Caiazzo}, {van Roestel}, {Vanderbosch}, {El-Badry},
  {Bellm}, {G{\"a}nsicke}, {Graham}, {Mahabal}, {Masci}, {Mr{\'o}z}, {Riddle},
  \& {Rusholme}}]{2023ApJ...945..141R}
{Rodriguez}, A.~C., {Kulkarni}, S.~R., {Prince}, T.~A., {et~al.} 2023, \apj,
  945, 141

\bibitem[{{Saxton} {et~al.}(2008){Saxton}, {Read}, {Esquej}, {Freyberg},
  {Altieri}, \& {Bermejo}}]{Saxton2008}
{Saxton}, R.~D., {Read}, A.~M., {Esquej}, P., {et~al.} 2008, \aap, 480, 611

\bibitem[{{Scargle}(1982)}]{Scargle1982}
{Scargle}, J.~D. 1982, \apj, 263, 835

\bibitem[{{Schlafly} \& {Finkbeiner}(2011)}]{Schlafly2011}
{Schlafly}, E.~F. \& {Finkbeiner}, D.~P. 2011, \apj, 737, 103

\bibitem[{{Schmidt} {et~al.}(2005){Schmidt}, {Szkody}, {Vanlandingham},
  {et~al.}}]{Schmidt2005LowAccretion}
{Schmidt}, G.~D., {Szkody}, P., {Vanlandingham}, K.~M., {et~al.} 2005, \apj,
  630, 1037

\bibitem[{Schreiber {et~al.}(2021)Schreiber, Belloni, G{\"a}nsicke, Parsons, \&
  Zorotovic}]{2021schreiber}
Schreiber, M.~R., Belloni, D., G{\"a}nsicke, B.~T., Parsons, S.~G., \&
  Zorotovic, M. 2021, Nature Astronomy, 5, 648

\bibitem[{Schreiber {et~al.}(2024)Schreiber, Belloni, \&
  Schwope}]{2024schreiber}
Schreiber, M.~R., Belloni, D., \& Schwope, A.~D. 2024, \aap, 682, L7, comment:
  Accepted for publication in Astronomy \& Astrophysics Letters

\bibitem[{{Schwope} {et~al.}(2024){Schwope}, {Knauff}, {Kurpas}, {Salvato},
  {Stelzer}, \& et~al.}]{Schwope2024erosita}
{Schwope}, A., {Knauff}, K., {Kurpas}, J., {et~al.} 2024, \aap, 690, A243

\bibitem[{Schwope(2025)}]{2025schwope}
Schwope, A.~D. 2025, \aap, 698, A106

\bibitem[{{Silber}(1992)}]{Silber1992}
{Silber}, A.~D. 1992, PhD thesis, Massachusetts Institute of Technology

\bibitem[{{Skrutskie} {et~al.}(2006){Skrutskie}, {Cutri}, {Stiening},
  {et~al.}}]{Skrutskie2006}
{Skrutskie}, M.~F., {Cutri}, R.~M., {Stiening}, R., {et~al.} 2006, \aj, 131,
  1163

\bibitem[{Storn \& Price(1997)}]{StornPrice1997}
Storn, R. \& Price, K. 1997, Journal of Global Optimization, 11, 341

\bibitem[{{Sun} {et~al.}(2021){Sun}, {Cheng}, {Ye}, \& et~al.}]{Sun2021}
{Sun}, Y., {Cheng}, Z., {Ye}, S., \& et~al. 2021, \apjs, 257, 65

\bibitem[{{Szkody} {et~al.}(2003){Szkody}, {Homer}, {Chen},
  {et~al.}}]{Szkody2003RareMagneticCVs}
{Szkody}, P., {Homer}, L., {Chen}, B., {et~al.} 2003, \apj, 583, 902

\bibitem[{Tout {et~al.}(2008)Tout, Wickramasinghe, Liebert, Ferrario, \&
  Pringle}]{2008tout}
Tout, C.~A., Wickramasinghe, D.~T., Liebert, J., Ferrario, L., \& Pringle,
  J.~E. 2008, \mnras, 387, 897

\bibitem[{{Vaeth} \& {Chanmugam}(1995)}]{Vaeth1995}
{Vaeth}, H.~M. \& {Chanmugam}, G. 1995, \apjs, 98, 295

\bibitem[{{van Roestel} {et~al.}(2025){van Roestel}, {Rodriguez}, {Szkody},
  {Brown}, {Caiazzo}, {Drake}, {El-Badry}, {Prince}, {Rich}, {Neill},
  {Vanderbosch}, {Bellm}, {Dekany}, {Feinstein}, {Graham}, {Groom}, {Helou},
  {Kulkarni}, {du Laz}, {Mahabal}, {Sharma}, {Sollerman}, \&
  {Wold}}]{2025A&A...696A.242V}
{van Roestel}, J., {Rodriguez}, A.~C., {Szkody}, P., {et~al.} 2025, \aap, 696,
  A242

\bibitem[{{VanderPlas}(2018)}]{VanderPlas2018}
{VanderPlas}, J.~T. 2018, \apjs, 236, 16

\bibitem[{Virtanen {et~al.}(2020)}]{Virtanen2020}
Virtanen, P. {et~al.} 2020, Nature Methods, 17, 261

\bibitem[{{Visvanathan} \& {Wickramasinghe}(1979)}]{1979Natur.281...47V}
{Visvanathan}, N. \& {Wickramasinghe}, D.~T. 1979, \nat, 281, 47

\bibitem[{Warner(1995)}]{1995warner}
Warner, B. 1995, Cambridge Astrophysics Series, Vol.~28, Cataclysmic Variable
  Stars (Cambridge: Cambridge University Press)

\bibitem[{{Westfold}(1959)}]{1959westfold}
{Westfold}, K.~C. 1959, \apj, 130, 241

\bibitem[{{Wickramasinghe}(1988)}]{1988prco.book....3W}
{Wickramasinghe}, D.~T. 1988, in Polarized Radiation of Circumstellar Origin,
  ed. G.~V. {Coyne}, A.~M. {Magalhaes}, A.~F. {Moffat}, R.~E.
  {Schulte-Ladbeck}, \& S.~{Tapia} (Vatican Observatory), 3--44

\bibitem[{{Wickramasinghe} \& {Ferrario}(2000)}]{wickramasinghe2000}
{Wickramasinghe}, D.~T. \& {Ferrario}, L. 2000, \pasp, 112, 873

\bibitem[{{Wickramasinghe} \& {Meggitt}(1985)}]{1985WM}
{Wickramasinghe}, D.~T. \& {Meggitt}, S.~M.~A. 1985, \mnras, 214, 605

\bibitem[{{Wickramasinghe} {et~al.}(1987){Wickramasinghe}, {Tuohy}, \&
  {Visvanathan}}]{wickramasinghe1987}
{Wickramasinghe}, D.~T., {Tuohy}, I.~R., \& {Visvanathan}, N. 1987, \apj, 318,
  326

\bibitem[{{Williams}(1980)}]{Williams1980}
{Williams}, R.~E. 1980, \apj, 235, 939

\bibitem[{{Wright} {et~al.}(2010){Wright}, {Eisenhardt}, {Mainzer},
  {et~al.}}]{Wright2010}
{Wright}, E.~L., {Eisenhardt}, P.~R.~M., {Mainzer}, A.~K., {et~al.} 2010, \aj,
  140, 1868

\end{thebibliography}
\end{document}